\documentclass[11pt]{iopart}
\usepackage{comment}
\usepackage{enumerate}
\usepackage{amssymb}
\expandafter\let\csname equation*\endcsname\relax
\expandafter\let\csname endequation*\endcsname\relax
\usepackage{amsmath}
\usepackage{braket}
\usepackage{dsfont}
\usepackage{graphicx}
\usepackage[usenames,dvipsnames]{color}
\usepackage{tensor}
\usepackage{empheq}
\usepackage{capt-of}
\usepackage[normalem]{ulem} % either use this (simple) or
\usepackage{soul} % 
\usepackage{MnSymbol}
\usepackage{cancel}

\usepackage[colorlinks,bookmarks=false,citecolor=NavyBlue,linkcolor=OliveGreen,urlcolor=blue]{hyperref}
\newcommand{\be}{\begin{equation}}
\newcommand{\ee}{\end{equation}}
\newcommand{\ba}{\begin{aligned}}
\newcommand{\ea}{\end{aligned}}
\newcommand{\bw}{\begin{widetext}}
\newcommand{\ew}{\end{widetext}}

\newcommand{\bea}{\begin{eqnarray}}
\newcommand{\eea}{\end{eqnarray}}

\def\doi{http://dx.doi.org/}

\begin{document}
\title{Entanglement negativity in a fermionic chain with dissipative defects: Exact results}
\author{Fabio Caceffo$^1$ and Vincenzo Alba$^1$}
\address{
$^1$Dipartimento di Fisica dell' Universit\`a di Pisa and INFN, Sezione di Pisa, I-56127 Pisa, Italy\\
%* vincenzo.alba@unipi.it
}

\begin{abstract}
	We investigate the dynamics of the fermionic logarithmic negativity 
	in a free-fermion chain with a localized loss, which acts as a dissipative 
	impurity. The chain is initially prepared in a generic 
	Fermi sea. In the standard hydrodynamic limit of 
	large subsystems and long times, with their ratio fixed, 
	the negativity between two subsystems is described by a simple formula, which depends only on 
	the effective absorption coefficient of the impurity. The negativity grows 
	linearly at short times, then saturating to a volume-law scaling. Physically, 
	this reflects the continuous production with time of entangling pairs of excitations 
	at the impurity site. Interestingly, the negativity is not 
	the same as the R\'enyi mutual information with R\'enyi index $1/2$, in contrast with the 
	case of unitary dynamics. 
	This  reflects the interplay between dissipative and unitary processes. 
	The negativity content of the entangling pairs is obtained in terms of an 
	effective two-state mixed density matrix for the subsystems. 
	Criticality in the initial Fermi sea is reflected 
	in the presence of logarithmic corrections. 
	The prefactor of the logarithmic scaling depends on 
	the loss rate, suggesting a nontrivial interplay between dissipation and 
	criticality. 
\end{abstract}

\maketitle

%##########################
\section{Introduction}

In recent years tremendous progress in cold-atom experiments~\cite{islam2015measuring,kaufman2016quantum,brydges2019probing,elben2020mixed,neven2021symmetry,vitale2022symmetry,elben2022the} and the rise of Noisy-Intermediate-Scale-Quantum~\cite{preskill2018quantum,preskill2022quantum} (NISQ) 
platforms provided us with the opportunity to test quantum mechanics with an 
unprecedented level of accuracy and versatility. 
The question how the macroscopic ``classical'' world emerges and interacts 
with the microscopic quantum one is one of the most fundamental research themes. 

\begin{figure}
	\begin{center}
		\includegraphics[width=0.9\linewidth]{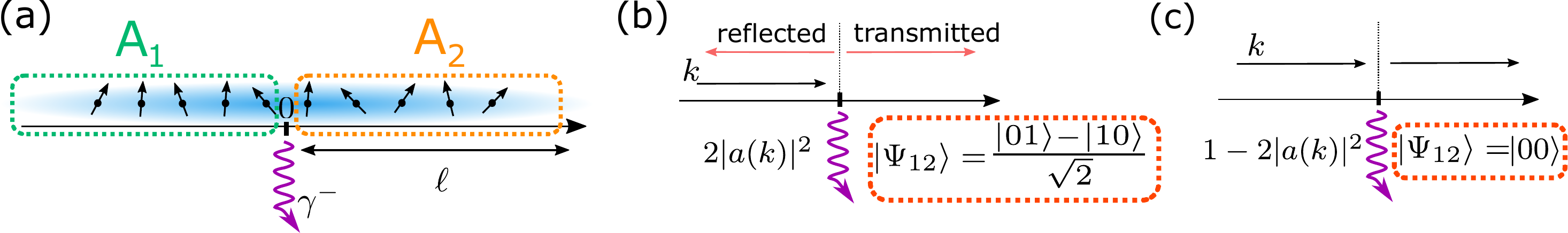}
		\caption{A free fermion chain subject to a localized fermion loss. (a) The 
			chain is prepared in the uniform Fermi sea with Fermi momentum $k_F$. 
			At the origin  fermions are removed incoherently at 
			a rate $\gamma^-$. We are interested in the fermionic logarithmic negativity 
			between two equal intervals $A_1$ and $A_2$ of length $\ell$ 
			embedded in an infinite chain and adjacent to the fermionic loss. (b) and (c) 
			Mechanism for entanglement generation via the lossy site. In (b) a fermion 
			with quasimomentum $k$ is absorbed at the origin with probability 
			$2|a(k)|^2$, with $|a(k)|^2$ the 
			absorption coefficient. The two subsystems are put in the maximally entangled 
			state $(|01\rangle-|10\rangle)/\sqrt{2}$, with $|0\rangle$ and $|1\rangle$ 
			normalized states. (c) The fermion is not affected by the loss with 
			probability $1-2|a(k)|^2$, implying that the two subsystems are in the product state $|00\rangle$. 
		}
		\label{fig:cartoon}
	\end{center}
\end{figure}

Here we address the question how entanglement, which is the distinctive feature of quantum 
mechanics, is affected by the interaction with the environment in \emph{open} quantum 
many-body systems. Specifically, we consider the paradigmatic setup of a 
uniform Fermi sea subject to the incoherent removal 
of fermions at the centre of the system. This localized fermionic loss can be 
viewed as a dissipative impurity (see Fig.~\ref{fig:cartoon}). 
We model the system-environment interaction with the Lindblad 
master equation~\cite{petruccione2002the,rossini2021coherent}. 
Recently, it has been shown that in the presence of \emph{global} dissipation 
modelled by a quadratic Lindblad equation~\cite{prosen2008third} it is possible to 
use the hydrodynamic framework to describe the dynamics of several entanglement-related 
quantities~\cite{alba2021spreading,carollo2022dissipative,alba2022hydrodynamics}, 
such as the von Neumann entropy and R\'enyi entropies, the mutual information, and the logarithmic 
negativity~\cite{alba2022logarithmic}. 
These results generalize a well-established 
quasiparticle picture~\cite{calabrese2005evolution,fagotti2008evolution,alba2017entanglement,alba2018entanglement,alba2021generalized} 
for entanglement spreading to free systems 
with quadratic dissipation~\cite{prosen2008third} \emph{\`a la } Lindblad.  
Crucially, in the presence of an environment  genuine quantum 
entanglement and spurious statistical correlations are deeply intertwined, since  
the global system is in a mixed state. Neither the von Neumann entropy  
nor the mutual information are proper measures of entanglement. Instead, the logarithmic 
negativity~\cite{vidal2002computable} or the fermionic logarithmic negativity~\cite{shapourian2019entanglement} 
can be used to sieve genuine entanglement even in mixed-state systems. 
Moreover, while quite generically global dissipation destroys entanglement, 
dissipative localized impurities, such as fermionic 
gain/loss, can induce robust entanglement production~\cite{alba2021unbounded}.

Free systems with localized losses provide a minimal theoretical laboratory to understand 
the interplay between quantum and statistical correlations underpinning   
the entanglement growth. 
Statistical correlations are due to the initial homogeneous Fermi sea 
being progressively depleted with time. Specifically, the absorption of 
fermions transforms the initial 
zero-entropy Fermi sea in a state with finite thermodynamic entropy. This finite 
thermodynamic entropy is reflected in a generic growth of the von Neumann entropy~\cite{alba2021unbounded}. 
At the same time, genuine entanglement production occurs. 
The mechanism at play is depicted in Fig.~\ref{fig:cartoon} (b). 
Each fermion in the initial state can scatter with the dissipative impurity. 
This leads to a continuous production in time 
of entangled pairs formed by the reflected and transmitted fermions. The  
entangled fermions propagate in the system 
entangling different spatial regions. A similar mechanism governs the entanglement dynamics 
in the presence of impurities in closed systems~\cite{eisler2012entanglement}. 
The production of genuine entanglement is captured by the negativity, whereas the von Neumann 
entropy is sensitive to both statistical and quantum correlations. 
Remarkably, in closed systems with defects the negativity becomes half of the R\'enyi mutual 
information with R\'enyi index $1/2$~\cite{gruber2020time,fraenkel2022extensive}, 
similar to what found in translation invariant setups~\cite{alba2019quantum}. 
%
%#############################################################
%#############################################################
%
As we are going to show, this is not the case in the presence of dissipative impurities.

Specifically, here we derive a formula describing the fermionic logarithmic negativity between two 
equal-length intervals in a free-fermion chain in the presence of localized 
losses (see Fig.~\ref{fig:cartoon} (a)). This allows us to precisely unravel how statistical correlations  trigger a 
linear growth of entanglement. 
Our formula holds in the 
standard hydrodynamic limit of large intervals and long times, with their ratio fixed. In this 
limit the negativity is described by a simple hydrodynamic formula, which depends on the absorption 
coefficient of the impurity~\cite{alba2022noninteracting}. The negativity 
grows linearly with time up to times proportional to the size of the intervals, then attaining a volume-scaling 
at infinite time. In contrast with the von Neumann entropy, the negativity is only sensitive to 
the entangled fermionic pairs produced at the impurity, as we verify by 
varying the geometry of the two intervals. Specifically, the negativity is proportional to the 
number of entangled pairs that are 
shared between the intervals, and it is not related to the R\'enyi mutual information, 
in contrast with the results in the unitary case~\cite{alba2019quantum,gruber2020time,fraenkel2022extensive}. 
Similar behaviour is observed for the case of global dissipation~\cite{alba2021unbounded}. 
The negativity content  of the quasiparticles 
can be derived from an effective two-state mixed density matrix for the intervals. 
Finally, going beyond the hydrodynamic limit, we numerically show that the negativity 
exhibits subleading terms that scale logarithmically with the intervals size. While these terms are 
reminiscent of the Fermi sea critical behaviour, their prefactor  
depends on the loss rate. This suggests a deep interplay between dissipation 
and criticality. 

The outline of the paper is as follows. We start in section~\ref{sec:model} with a review of the 
free-fermion chain with a localized loss. In particular, we provide the result for the fermionic 
correlation function~\eqref{eq:Gxy-ansatz}, which is necessary to obtain the negativity. 
In section~\ref{sec:negativity} we introduce the negativity, and its calculation in free-fermion 
systems. In section~\ref{sec:main-result} we present without derivation our main result (see formula~\eqref{eq:main}). 
We compare our results with numerical data in section~\ref{sec:numerics}, 
discussing also some interesting logarithmic corrections to the hydrodynamic 
limit. Finally, we conclude in section~\ref{sec:conclusions}. In~\ref{sec:app-1} we 
report the derivation of the results of section~\ref{sec:main-result}, in particular of 
formula~\eqref{eq:main}. In~\ref{sec:app-2} we discuss the tripartition of the system with 
$A_1$ and $A_2$ on the same side of the impurity, showing that the negativity does not 
grow linearly with time. Finally, in~\ref{sec:app-3} we report the calculation of the 
logarithmic negativity in the low-density limit, in which there is only one fermion in the system. 
%#############################################################
%

%##########################
\section{Preliminaries: Tight-binding chain with localized losses}
\label{sec:model}

Here  we focus on the tight-binding chain with localized losses. The Hamiltonian 
describes a system of free fermions hopping coherently between nearest-neighbour 
sites, and it is defined as 
\begin{equation}
\label{eq:ham}
H=-J\sum_{x=-\infty}^\infty(c^\dagger_xc_{x+1}+c^{\dagger}_{x+1}c_x),
\end{equation}
where $c^\dagger_x$ and $c_x$ are standard fermion creation and annihilation 
operators, and $J$ is the hopping strength. Here we fix $J=1$. We consider an 
infinite chain. The model is diagonalized by going to Fourier space, defining 
new fermionic operators $b_k$ as 
\begin{equation}
b_k:=\sum_{x=-\infty}^\infty e^{-ikx}c_x,\quad c_x=\int_{-\pi}^\pi\frac{dk}{2\pi}e^{ikx}b_k. 
\end{equation}
In Fourier space Eq.~\eqref{eq:ham} becomes diagonal as  
\begin{equation}
H=\int_{-\pi}^\pi\frac{dk}{2\pi}\varepsilon_k b^\dagger_k b_k. 
\end{equation}
where the single-particle dispersion $\varepsilon_k$ and the group velocity 
of the fermions $v_k$ read as 
\begin{equation}
\label{eq:vk}
\varepsilon_k:=-2\cos(k),\quad v_k:=\frac{\partial \varepsilon_k}{\partial k}=2\sin(k). 
\end{equation}
The Hamiltonian~\eqref{eq:ham} commutes with total number of fermions. The ground state 
of the chain is obtained by filling all the single-particle levels up to the Fermi 
momentum $k_F$, i.e., with $k\in[-k_F,k_F]$. Here $k_F$ is related to the fermion density 
$n_f$ as $n_f=k_F/\pi$. For $n_f=1$ the ground state of~\eqref{eq:ham} becomes a trivial  
band insulator. At $n_f<1$, i.e., for $|k_F|<\pi$ the ground state is a critical state with 
power-law decaying correlations. 

In the presence of localized losses the evolution of the system can be modelled by 
the Lindblad equation~\cite{petruccione2002the}. The evolution of the system density matrix 
$\rho_t$ is described by 
\begin{equation}
\label{eq:lin}
\frac{d\rho_t}{dt}={\mathcal L}(\rho_t)=-i[H,\rho_t]+\gamma^- c_0\rho_t c_0^\dagger-\frac{1}{2}\{\gamma^-
c_0^\dagger c_0,\rho_t\}, 
\end{equation}
where $\gamma^-$ is the rate at which fermions are removed at the origin, i.e., the loss 
rate. 

The fermionic correlation function $G_{x,y}$ is a central object to determine entanglement 
properties of the chain~\cite{peschel2009reduced}. The correlation matrix is defined as 
\begin{equation}
\label{eq:Gxy}
G_{x,y}:=\mathrm{Tr}(c^\dagger_x c_y\rho_t). 
\end{equation}
The Lindblad equation~\eqref{eq:lin} can be used to obtain the dynamics of the $G_{x,y}$ as 
\begin{equation}
\label{eq:G-lin}
\frac{dG_{x,y}}{dt}=i(G_{x+1,y}+G_{x-1,y}-G_{x,y+1}-G_{x,y-1})-
\frac{\gamma^-}{2}(\delta_{x,0}+\delta_{y,0})G_{x,y}. 
\end{equation}
Eq.~\eqref{eq:G-lin} is obtained by observing that the dynamics of a generic 
observable $\hat {\mathcal O}$ from~\eqref{eq:lin} is obtained as 
\begin{equation}
	\frac{d\hat {\mathcal O}}{dt}={\mathcal L}^\dagger(\hat {\mathcal O}), 
\end{equation}
where the Liouvillian operator ${\mathcal L}$ is defined in~\eqref{eq:lin}. 
Eq.~\eqref{eq:G-lin} is obtained by choosing $\hat{\mathcal O}=c_x^\dagger c_y$, 
by taking the expectation value $\langle\cdot\rangle=\mathrm{Tr}(\cdot \rho)$, and 
using Wick theorem. 

If the initial state is a Fermi sea with Fermi momentum $k_F$, Eq.~\eqref{eq:G-lin} 
can be solved by using a product ansatz for $G_{x,y}$ as 
\begin{equation}
\label{eq:Gxy-ansatz}
G_{x,y}(t)=\int_{-k_F}^{k_F}\frac{dk}{2\pi}S_{k,x}\bar S_{k,y}, 
\end{equation}
with $S_{k,x}$ to be determined, and the bar in $\bar S_{k,x}$ denoting 
complex conjugation. For generic $x,t$ the final result is cumbersome, although it 
can be obtained in some cases (see, for instance, Ref.~\cite{krapivsky-2019} for the 
case with $k_F=\pi$). However, in the hydrodynamic limit $x,y,t\to\infty$ with $x/t,y/t$ fixed and 
$|x-y|/t\to0$, $S_{k,x}$ takes the simple form as~\cite{alba2022noninteracting}  
\begin{equation}
\label{eq:S}
S_{k,x}=e^{ik x}+\Theta(|v_k|t-|x|) r(k) e^{i|kx|}. 
\end{equation}
Eq.~\eqref{eq:S} suggests that in the hydrodynamic limit the dissipative site acts like 
an impurity. In~\eqref{eq:S}, $v_k$ is the fermion group velocity~\eqref{eq:vk}, $\Theta(z)$ 
the Heaviside step function, and $r(k)$ is the reflection amplitude. Here reflection and 
transmission amplitudes $r(k),t(k)$ are defined as~\cite{alba2022noninteracting} 
\begin{equation}
\label{eq:r-coeff}
r(k):=-\frac{\gamma^-}{2}\frac{1}{\frac{\gamma^-}{2}+|v_k|},\quad 
t(k):=\frac{|v_k|}{\frac{\gamma^-}{2}+|v_k|}.  
\end{equation}
A similar expression as Eq.~\eqref{eq:S} is obtained considering the one-dimensional 
scattering problem of a plane wave on a delta potential with imaginary strength~\cite{burke2020non}, i.e., 
$V(x)=-i\gamma/2 \delta(x)$. Now, the solution of the Schr\"odinger equation has the form 
$e^{ikx}+r'(k)e^{-ikx}$, with 
\begin{equation}
	r'(k)=-\frac{\gamma}{2}\frac{1}{\frac{\gamma}{2}+\frac{\hbar^2 k}{m}}. 
\end{equation}
Clearly, $r'(k)$ becomes the same as $r(k)$ in~\eqref{eq:r-coeff} after redefining 
$\hbar^2 k/m\to |v_k|$. 

Since the dynamics is not unitary, one has that $|r(k)|^2+|t(k)|^2\ne 1$. One 
can define the absorption coefficient for the fermions $a(k)$ as 
\begin{equation}
\label{eq:ak}
|a(k)|^2:=1-|r|^2-|t|^2=\gamma^-\frac{|v_k|}{\left(\frac{\gamma^-}{2}+|v_k|\right)^2}. 
\end{equation}
Eq.~\eqref{eq:S} can be rewritten as 
\begin{equation}
\label{eq:S-1}
S_{k,x}=e^{ikx} + r(k)e^{i|kx|}\int_{-\infty}^{\infty}\frac{dq}{2\pi i} 
\frac{e^{i(|v_k|t - |x|)q}}{q-i0^+},  
\end{equation}
where the integral in the second term is used to represent the Heaviside step in~\eqref{eq:S}, 
and $i0^+$ is an infinitesimal shift along the positive imaginary axis.
%#############################################################
%

%##########################
\section{Entanglement negativity for fermionic systems: Definitions}
\label{sec:negativity}

We are interested to quantify the entanglement between two adjacent regions $A_1$ and $A_2$ 
of equal length $\ell$ adjacent to the dissipative impurity (see Fig.~\ref{fig:cartoon} (a)). 
Due to the presence of dissipation, the chain is not in a global pure state. This implies that 
neither the R\'enyi entropies (and von Neumann entropy) nor the mutual information are proper 
entanglement measures of the 
entanglement between $A_1$ and $A_2$. In this situation, only 
the logarithmic negativity provides a proper entanglement measure. 
Here we consider the \emph{fermionic} logarithmic negativity~\cite{shapourian2019entanglement}. 
In contrast with the standard definition of negativity~\cite{vidal2002computable}, the fermionic one 
can be effectively 
computed from the fermionic two-point correlation function (cf.~\eqref{eq:Gxy}). 
To introduce the negativity, one first defines the restricted correlation matrix $G_A$ as 
\begin{equation}
G_A:=G_{x,y},\quad\mathrm{with}\,\, x,y\in A, 
\end{equation}
where $G_{x,y}$ is given in~\eqref{eq:Gxy}, and the matrix $\Gamma_A$ as
\begin{equation}
	\label{eq:gamma-a}
	\Gamma_A:=\mathds{1} - 2 G_A.
\end{equation}
To proceed, one defines the two matrices $\Gamma^\pm_A$ as 
\begin{equation}
\label{eq:gpm}
\Gamma^\pm_A=\left(
\begin{array}{cc}
-\Gamma_A^{11} & \pm i \Gamma_A^{12}\\
\pm i \Gamma_A^{21} & \Gamma_A^{22}
\end{array}
\right),\quad \Gamma_A^{ij}:=\Gamma_{x,y}\,\,\mathrm{with}\,\, x\in A_i,\,\,y\in A_j. 
\end{equation}
The fermionic logarithmic negativity is defined as~\cite{shapourian2017many,shapourian2017partial,shapourian2019entanglement,shapourian2019finite}. 
\begin{equation}
\label{eq:neg-def}
{\cal E}:=\mathrm{Tr}\ln\left[\sqrt{\frac{\mathds{1}-\Xi_A}{2}}
+\sqrt{\frac{\mathds{1}+\Xi_A}{2}}\right]+\mathrm{Tr}\ln\left[
\sqrt{G_A^2+(\mathds{1}-G_A)^2}\right],  
\end{equation}
where $\Xi_A$ is defined as 
\begin{equation}
\label{eq:xi-def}
\Xi_A:=P^{-1}(\Gamma^+_A+\Gamma^-_A), \qquad P:=\mathds{1}+\Gamma^+_A\Gamma^-_A. 
\end{equation}
In~\eqref{eq:xi-def},  $\Gamma^\pm_A$ are defined in~\eqref{eq:gpm}, and $\mathds{1}$ is the 
$2\ell\times2\ell$ identity matrix. 

For the following it is useful to compare the fermionic negativity with the original 
negativity~\cite{vidal2002computable}. 
The logarithmic negativity is defined in terms of the so-called partially transposed reduced 
density matrix. Given the reduced density matrix $\rho_A$ for 
$A_1\cup A_2$ (see Fig.~\ref{fig:cartoon}), we define the partial transpose $\rho_A^{\mathrm{T_2}}$ as 
\begin{equation}
\langle e^{(1)}_i\otimes e^{(2)}_j|\rho_A^{\mathrm{T_2}}|e_k^{(1)}\otimes e_l^{(2)}\rangle:=
\langle e^{(1)}_i\otimes e^{(2)}_l|\rho_A|e_k^{(1)}\otimes e_j^{(2)}\rangle, 
\end{equation}
where we denoted with $e_j^{(1)}$ and $e_j^{(2)}$ two bases for the Hilbert space of $A_1$ and $A_2$, respectively. 
For non-separable states, the partial transpose is not, in general, a positive-definite matrix, in contrast with $\rho_A$. The 
negative eigenvalues of $\rho_A^{\mathrm{T_2}}$ quantify the entanglement shared between the two 
intervals. The logarithmic negativity is defined as 
\begin{equation}
\label{eq:neg-or}
{\cal E}:=\ln(\mathrm{Tr}|\rho_A^\mathrm{T_2}|). 
\end{equation}
Unfortunately, calculating the logarithmic negativity~\cite{eisert1999a,vidal2002computable,plenio2005logarithmic,wichterich2009scaling,calabrese2012entanglement}  
in quantum-many-body systems is in general a challenging task~\cite{eisler2015on,herzog2016estimation}, except for 
free-boson systems~\cite{audenaert2002entanglement}. 
For free fermions the partial transpose cannot be written as a Gaussian fermionic operator, implying that its spectrum, and hence the 
negativity, cannot be computed effectively~\cite{eisler2015on}. 

Extracting the salient physical information, such as the scaling behaviour, of the negativity is 
in general challenging task. Still, several results are available.  
For equilibrium systems described by Conformal Field Theory, the scaling of the ground-state logarithmic 
negativity is known analytically~\cite{calabrese2012entanglement,calabrese2013entanglement,shapourian2017partial}, 
also at finite-temperature~\cite{calabrese2013entanglement,shapourian2019finite}. The behaviour of the 
negativity in massive quantum field theory was also investigated~\cite{fournier2016universal,hoogeveen2015entanglement}. For CFT systems, 
scaling properties of the full spectrum of the partial transpose 
are known~\cite{ruggiero2016negativity,shapourian2019twisted}. Interestingly, both the standard and 
the fermionic logarithmic negativity exhibit logarithmic scaling in random singlet 
phases of matter~\cite{ruggiero2016entanglement,shapourian2017partial}. 
The behaviour of the logarithmic negativity at finite-temperature criticality in quantum models has been 
investigated~\cite{wald2020entanglement,lu2020entanglement,wu2020entanglement}. 
Interestingly, the dynamics of negativity 
in integrable systems after a quantum quench~\cite{calabrese2016introduction} 
can be described within the quasiparticle picture~\cite{coser2014entanglement}. 
In the hydrodynamic regime of validity of the quasiparticle picture, both the fermionic and the standard negativity become 
half of the R\'enyi mutual information~\cite{alba2019quantum} with R\'enyi index $1/2$. 
This surprising result has been verified in free-fermion and free-boson models~\cite{alba2019quantum}, 
in free-fermion chains in the presence of defects~\cite{gruber2020time}, and in  
Conformal Field Theory~\cite{wen2015entanglement,kudler2020correlation,kudler2021the}. A rigorous 
derivation is also possible  in minimal models of interacting quantum many-body systems, such as dual 
unitary circuits~\cite{klobas2021exact,klobas2021entanglement}. Besides the negativity, 
the moments of the partial transpose, or the $PPT-n$ conditions, can be employed to quantify 
entanglement in many-body systems. The $PPT-n$ conditions can be also 
measured experimentally~\cite{elben2020mixed}, and admit a quasiparticle picture 
interpretation~\cite{murciano2022quench}. Some results are also available for disordered 
out-of-equilibrium systems~\cite{ruggiero2022quantum}. 

In the context of dissipative quantum many-body systems, 
the scaling of the fermionic logarithmic negativity has been investigated in the weakly-dissipative 
limit for free-fermions with gain and loss dissipation~\cite{alba2022logarithmic}. While a 
hydrodynamic description of the negativity is possible, the negativity content of the quasiparticles has no 
straightforward interpretation in terms of thermodynamic quantities. In particular, the relationship 
between negativity and mutual information established in Ref.~\cite{alba2019quantum} is violated. 
Very recently, the negativity has been studied in the context of free-fermions undergoing continuous 
monitoring~\cite{carollo2022entangled,turkeshi2022enhanced}.
%#############################################################
%

%##########################
\section{Logarithmic negativity in the presence of localized losses: Main result}
\label{sec:main-result} 

Here we outline our main result.  Let us consider the setup depicted in Fig.~\ref{fig:cartoon}: 
Two equal-length intervals  $A_1$ and $A_2$ are placed next to the dissipative impurity in an infinite chain.  
In the hydrodynamic limit $t,\ell\to\infty$ with the ratio $t/\ell$ fixed, and for generic 
loss rate $\gamma^-$, the fermionic logarithmic negativity between $A_1$ and $A_2$ is given as 
\begin{equation}
\label{eq:main}
{\cal E}=\frac{\ell}{2}
\int_{-k_F}^{k_F} \frac{dk}{2\pi} \min\left(|v_k|t/\ell,1\right)  e(1-2|a(k)|^2)
\end{equation}
with $v_k$ the fermion velocity~\eqref{eq:vk}, and $e(x)$ defined as  
\begin{equation}
\label{eq:neg-den}
e(x):=\ln\left(1-x+\sqrt{x^2+(1-x)^2}\;\right). 
\end{equation}
The negativity~\eqref{eq:main} depends only on the absorption coefficient 
$|a(k)|^2$ of the impurity (cf.~\eqref{eq:ak}). From Eq.~\eqref{eq:main}  one has 
that ${\cal E}$ grows linearly with time for $t<\ell/v_{\max}$, where $v_\mathrm{max}$ 
(cf.~\eqref{eq:vk}) is the maximum velocity for the fermions. Notice that 
$v_{\max}$ depends on the initial state via $k_F$. At asymptotically long times, 
${\cal E}$ saturates, and it exhibits a volume-law scaling ${\cal E}\sim\ell$. 
The physical interpretation of the growth of the negativity is as follows. 
As anticipated, the localized loss acts as a dissipative 
impurity. The fermions in the initial state scatter on the impurity 
giving rise to a superposition between transmitted and  reflected fermions, 
which form entangled pairs (see Fig.~\ref{fig:cartoon} (b)). 
The propagation of these entangled pairs in the bulk of the system  
gives rise to the entanglement between the two intervals. Specifically, 
if the reflected fermion is in $A_1$ and the transmitted is in $A_2$ they contribute 
to the negativity between them. 
The min function in~\eqref{eq:main} counts the number of entangled pairs thare are shared between 
the two subsystems. Finally, as there is a finite 
density of fermions in the initial state, there is a continuous  production in time of 
entangled pairs, which sustain the linear growth of ${\cal E}$. 
A similar mechanism explains the behaviour  of the entanglement entropy and 
of the negativity in free-fermion systems undergoing unitary dynamics in the presence of  
defects~\cite{eisler2012entanglement,collura2013entanglement,gruber2020time,gamayun2021nonequilibrium,gamayun2020fredholm,fraenkel2022extensive,capizzi2022entanglement}. More generically, the propagation of entangled pairs of 
quasiparticle excitations is the underlying mechanism for the entanglement 
dynamics in integrable systems~\cite{calabrese2005evolution,fagotti2008evolution,alba2017entanglement,alba2018entanglementdynamics} (see, however, 
Ref.~\cite{alba2017quench} and Ref.~\cite{bertini2022growth} for recent developments), also in the presence of quadratic Lindblad dissipation~\cite{alba2021spreading,carollo2022dissipative}. 
The derivation of~\eqref{eq:main} is technically involved, although 
the main tool is the multi-dimensional stationary phase approximation~\cite{wong}, 
together with the results of Ref.~\cite{alba2021unbounded}.
We report the derivation of~\eqref{eq:main} in~\ref{sec:app-1}.  

Instead, we now discuss the physical interpretation of the negativity content $e(k)$ (cf.~\eqref{eq:neg-den}) 
of the entangled pairs. 
Crucially, $e(k)$ is understood in terms of a two-level system for the subsystems. 
Let us introduce a fictitious state  $|s_1s_2\rangle$ 
for $A_1\cup A_2$. Here $s_i$ can take the values $0,1$. 
We can write the following density matrix for $A_1\cup A_2$ as 
\begin{equation}
\label{eq:rho12}
\rho_{A_1\cup A_2}=(1-2|a(k)|^2)|00\rangle\langle00|+|a(k)|^2
\left(|01\rangle-|10\rangle\right)\left(\langle01|-\langle10|\right). 
\end{equation}
The interpretation of~\eqref{eq:rho12} is that at the beginning the system is in the unperturbed product state $|00\rangle$ and the subsystems are not entangled with each other.
The presence of the dissipation, then, drives the system in the antisymmetric maximally entangled state $(|01\rangle - |10\rangle)/\sqrt{2}$ 
with probability $2|a(k)|^2$. The effective density matrix~\eqref{eq:rho12} and the mechanism by which fermion 
absorption leads to entanglement production can be also understood 
by considering the low-density limit in which there is only one fermion in the chain (see~\ref{sec:app-3}). 
The partially transposed reduced density matrix 
with respect to $A_2$ is obtained from~\eqref{eq:rho12} as
\begin{equation}
\label{eq:rhoT2}
\rho_{A_1\cup A_2}^{\mathrm{T_2}}=
(1-2|a(k)|^2)|00\rangle\langle00|+|a(k)|^2\left(|10\rangle\langle10|+|01\rangle\langle 01|
-|00\rangle\langle11|-|11\rangle\langle00|\right). 
\end{equation}
By calculating the negativity~\eqref{eq:neg-or} 
from~\eqref{eq:rhoT2} one obtains~\eqref{eq:neg-den} with $x=1-2|a(k)|^2$. 
Notice that here we are considering the original negativity~\eqref{eq:neg-or}, and 
not the fermionic one~\eqref{eq:neg-def}. Interestingly, the fact 
that from~\eqref{eq:rhoT2} we recover the $e(k)$ for the fermionic negativity 
could suggest that in the hydrodynamic limit the two definitions  
of the negativity are equivalent. This happens, for instance, if the dynamics is unitary~\cite{alba2019quantum}. However, 
checking this prediction is challenging because of the serious limitations of analytical and 
numerical methods to compute the negativity~\eqref{eq:neg-or} in out-of-equilibrium 
fermionic systems. 
As it is clear from~\eqref{eq:main}, the equality between R\'enyi mutual information and 
negativity established in~\cite{alba2019quantum} does not hold in the presence 
of dissipative impurities, although the fermionic negativity and the original one 
could become the same in the hydrodynamic limit. We should also mention that 
the functional form of $e(k)$ is reminiscent of the negativity between two intervals 
in a chain with a single fermion (see~\ref{sec:app-3} for the result). 
%
%#############################################################
\begin{figure}
	\centering
	\includegraphics[width=0.65\linewidth]{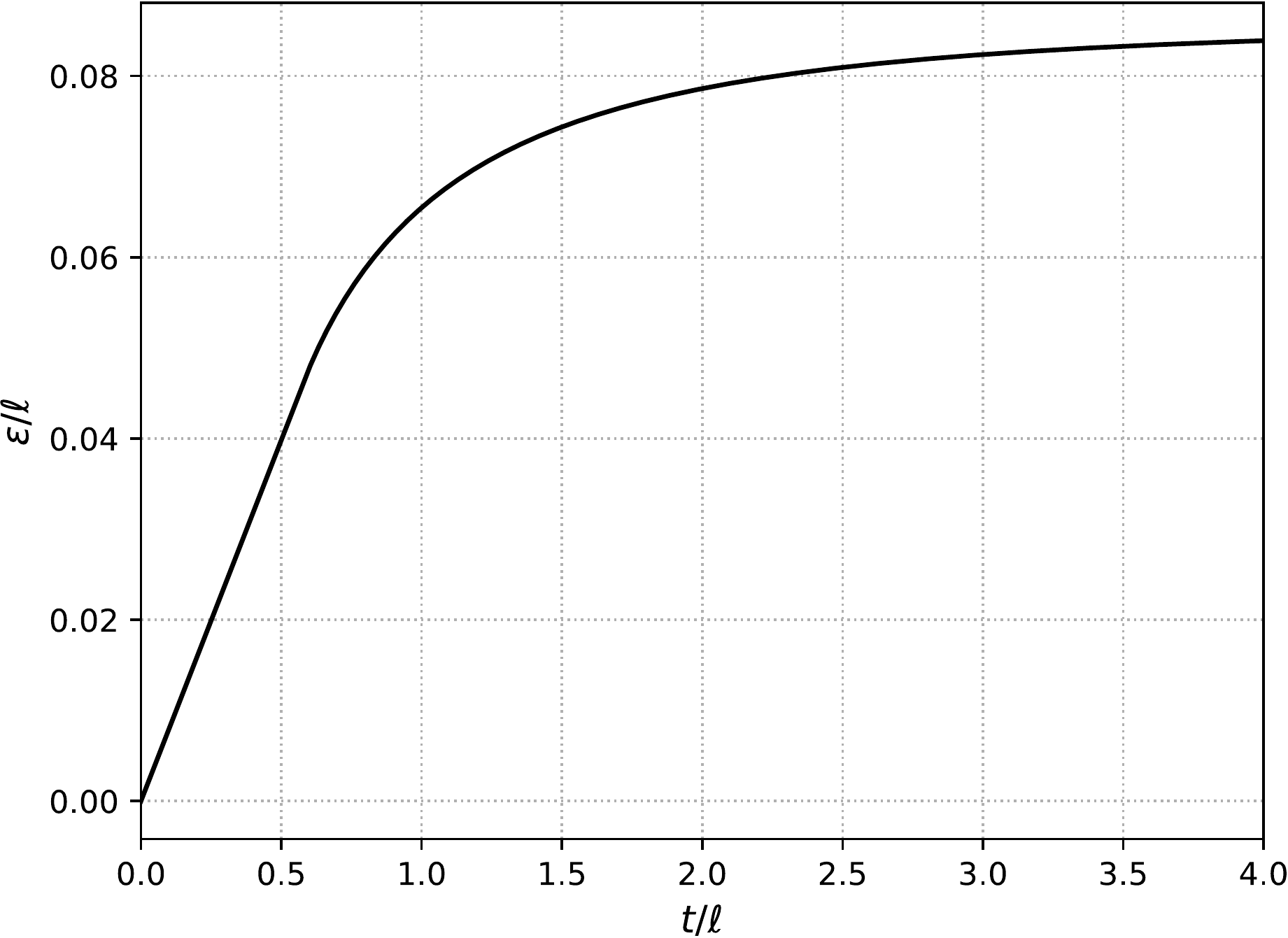}
	\caption{ Typical behavior of the fermionic negativity ${\cal E}$ in the free-fermion 
		chain with localized losses. The curve is the analytic result (cf.~\eqref{eq:main}) for a chain prepared 
		in a Fermi sea with $k_F=\pi/3$ and loss rate $\gamma^-=1$. In the figure we show 
		${\cal E}/\ell$ versus $t/\ell$, with $\ell$ the size of the two subsystems 
		(see Fig.~\ref{fig:cartoon}). Notice the linear increase up to $t/\ell=v_\mathrm{max}$, 
		followed by a saturating behavior for $t/\ell \to\infty$. 
	}
	\label{fig:negvstl}
\end{figure}
%#############################################################
%
Eq.~\eqref{eq:main} holds only for two adjacent intervals (see Fig.~\ref{fig:cartoon}), 
although it can be straightforwardly generalized to different geometries. Indeed, the function $e(k)$ remains 
the same, whereas the function $\min(|v(k)t|/\ell,t)$ has to be modified to account for the different 
kinematics of the quasiparticles. Importantly, for the 
geometry with the two intervals next to the impurity as in Fig.~\ref{fig:cartoon}, both the negativity and the von Neumann entropy 
grow with time~\cite{alba2021unbounded}. On the other hand, while the von Neumann entropy is sensitive to both 
statistical correlations and entanglement, the negativity is sensitive to entanglement only. This can be 
verified by considering the geometry with both intervals on the same side of the impurity. This 
is discussed in~\ref{sec:app-2}. 
Since the reflected and transmitted fermions are created at the impurity site and travel in 
opposite directions, they cannot be shared between the two intervals, implying that no entanglement 
is produced. While in this situation the von Neumann entropy is still nonzero~\cite{alba2021unbounded}, 
i.e., it fails to distinguish between classical and quantum correlations, the negativity vanishes 
in the scaling limit, as we prove in~\ref{sec:app-2}. 

\begin{figure}
	\centering
	\includegraphics[width=0.65\linewidth]{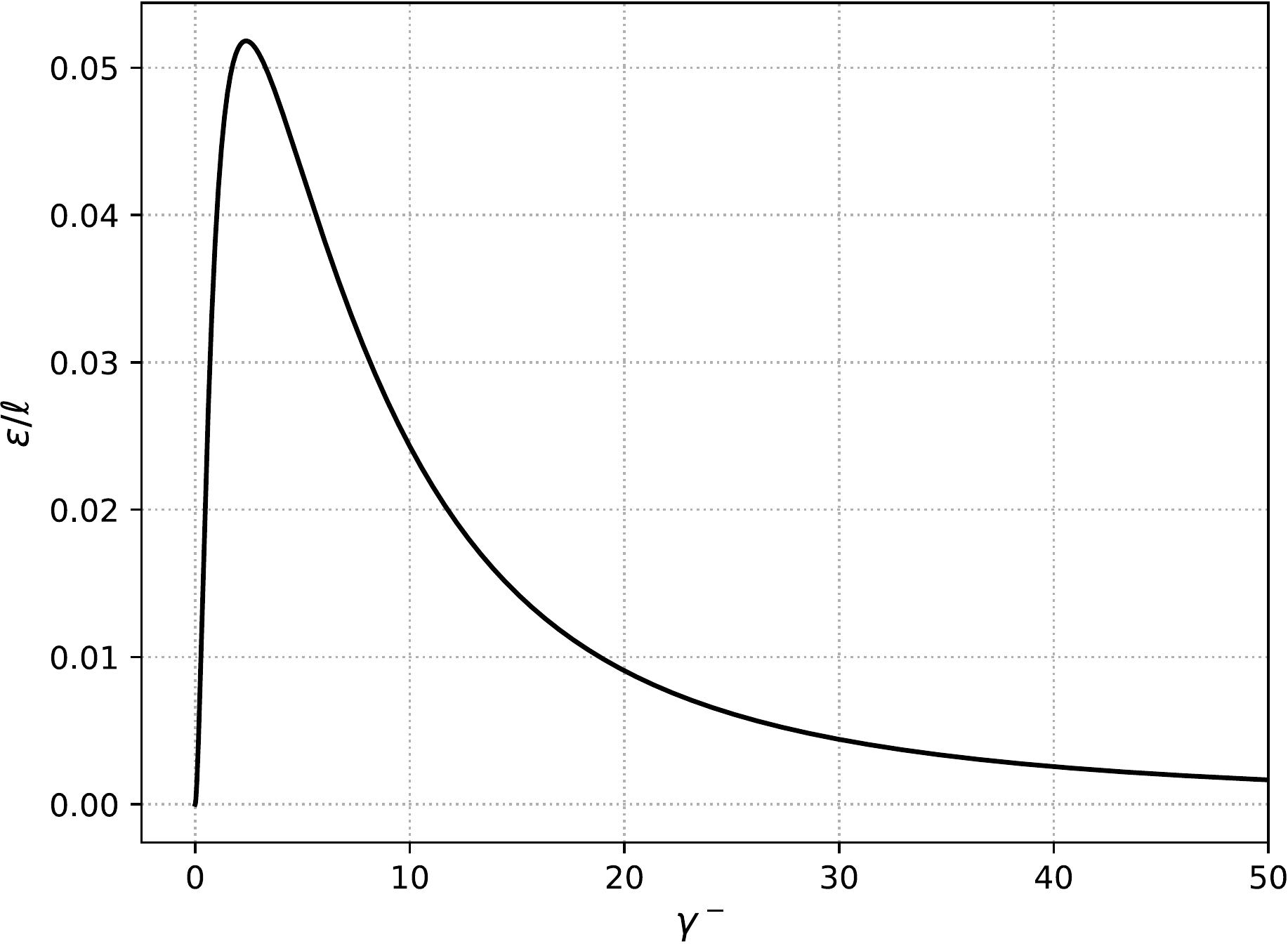}
	\caption{Logarithmic negativity ${\cal E}/\ell$ as a function of the loss rate $\gamma^-$ at fixed  
		$t/\ell=0.5$ and $k_F=\pi/3$. The figure shows the analytic result in the hydrodynamic 
		limit $t,\ell\to\infty$ with $t/\ell$ fixed (cf.~\eqref{eq:main}). Note the vanishing behavior 
		for $\gamma^-\to\infty$. 
	}
	\label{fig:negvsgamma}
\end{figure}

The generic behaviour of~\eqref{eq:main} as a function of time is illustrated in Fig.~\ref{fig:negvstl}. 
In the figure we show results for $k_F=\pi/3$. The setup is the same as in 
Fig.~\ref{fig:cartoon} (a). To highlight the scaling behaviour we plot 
${\cal E}/\ell$ versus $t/\ell$. As discussed above, ${\cal E}/\ell$ grows linearly 
with time up to $t^*=1/v_\mathrm{max}$. Here $v_\mathrm{max}=v_{\pi/3}<1$ (cf.~\eqref{eq:vk}). 
For $t\gg t^*$ the negativity saturates at asymptotically long times. It is  interesting to investigate 
the behaviour of the negativity as a function of the loss rate $\gamma^-$. This is discussed 
in Fig.~\ref{fig:negvsgamma}. We plot ${\cal E}/\ell$ versus $\gamma^-$ at 
fixed $t/\ell=1/2$. The negativity exhibits a maximum at intermediate times, then it decreases, 
vanishing in the limit $\gamma^-\to\infty$. A similar behaviour is observed for the von Neumann 
entropy~\cite{alba2021unbounded}. The vanishing behaviour at $\gamma^-\to\infty$ is a manifestation 
of the quantum Zeno effect~\cite{degasperis1974does,misra1977the,facchi2002quantum}. Precisely, 
in the limit $\gamma^-\to\infty$ the absorption and the transmission coefficients 
vanish (cf.~\eqref{eq:r-coeff} and~\eqref{eq:ak}), whereas the reflection coefficient 
$|r(k)|^2$ goes to one. This means that at large $\gamma^-$ the 
two halves of the chain are effectively decoupled. 

%
%#############################################################
%#############################################################
%

%##################################
\section{Numerical checks \& logarithmic corrections} 
\label{sec:numerics}

We now discuss numerical benchmarks of our main formula~\eqref{eq:main}, comparing it against 
exact numerical data. The data are obtained by solving numerically~\eqref{eq:G-lin}. The obtained  
fermionic correlator $G_{x,y}$ is then used to compute the negativity (see section~\ref{sec:negativity}). 
Eq.~\eqref{eq:G-lin} is solved for a finite chain of length $L$, with open boundary conditions. 
In our simulations we ensure that $t\ll L$ to avoid revival effects due to open boundary conditions. 

%
%#############################################################
\begin{figure}[t]
	\centering
	\includegraphics[width=0.65 \linewidth]{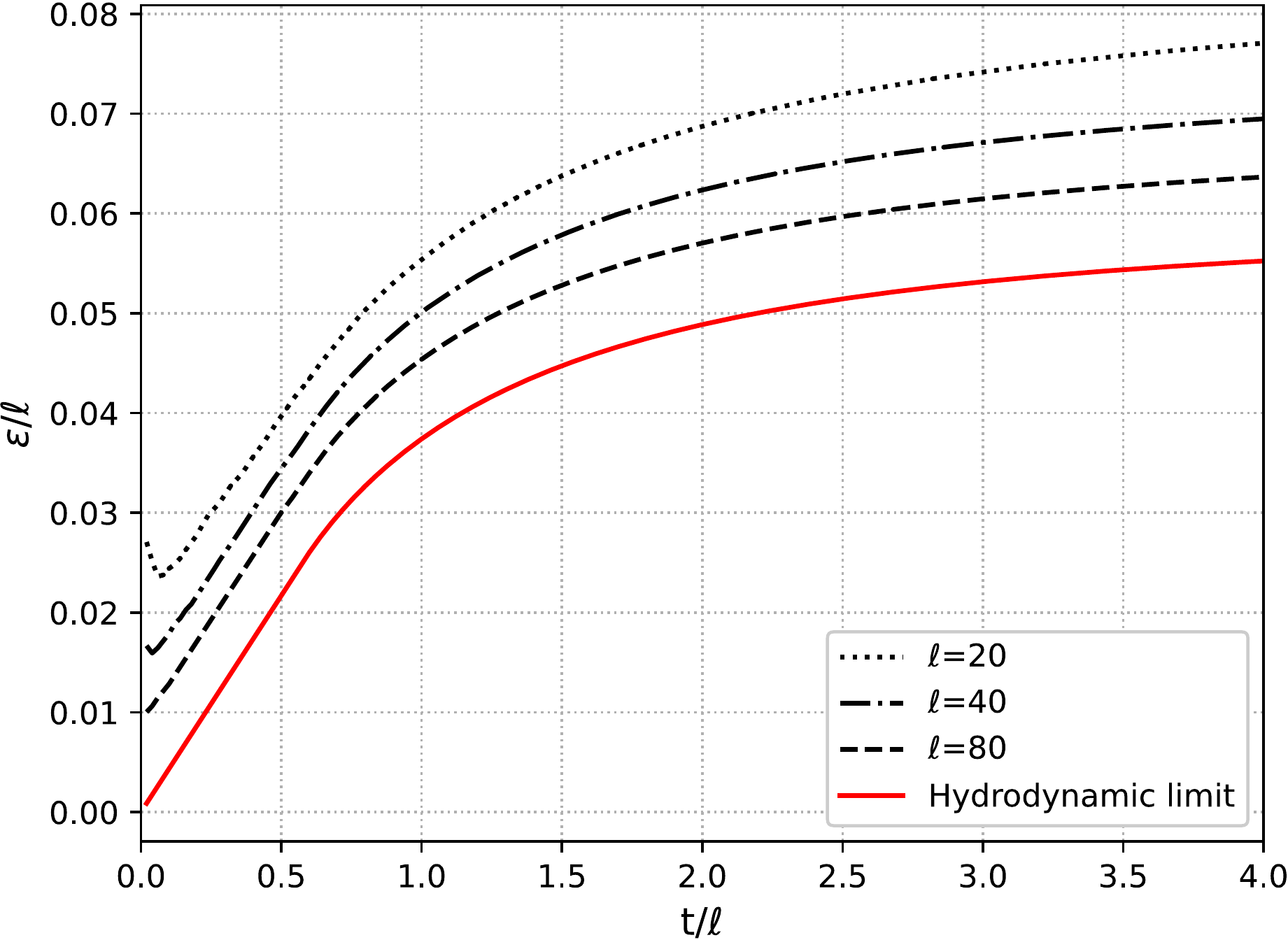}
	\caption{Logarithmic negativity between two equal intervals in a free-fermion chain 
		with a localized loss. Comparison between the theoretical prediction~\eqref{eq:main} in 
		the hydrodynamic limit (red continuous line) and numerical results for finite interval size 
		$\ell$ up to $\ell=80$. Data are for fixed loss rate $\gamma^-=0.5$. The chain is initially prepared in a 
		Fermi sea with $k_F=\pi/3$. Upon increasing $\ell$ the data approach the result in the 
		hydrodynamic limit. 
	}
	\label{fig:conv_full}
\end{figure}
%#############################################################
%
In Fig.~\ref{fig:conv_full} we show data for ${\cal E}/\ell$ plotted versus $t/\ell$ for the 
system with $\gamma^-=0.5$  and $k_F=\pi/3$. We show the theory prediction~\eqref{eq:main} (continuous line) and 
exact numerical  results for 
subsystem sizes $\ell=20,40,80$. Clearly, the finite-size data exhibit noticeable deviations from the 
hydrodynamic limit. Still, upon increasing $\ell$ the data  approach~\eqref{eq:main}. 
%
%#############################################################
\begin{figure}[t]
	\centering
	\includegraphics[width=0.65 \linewidth]{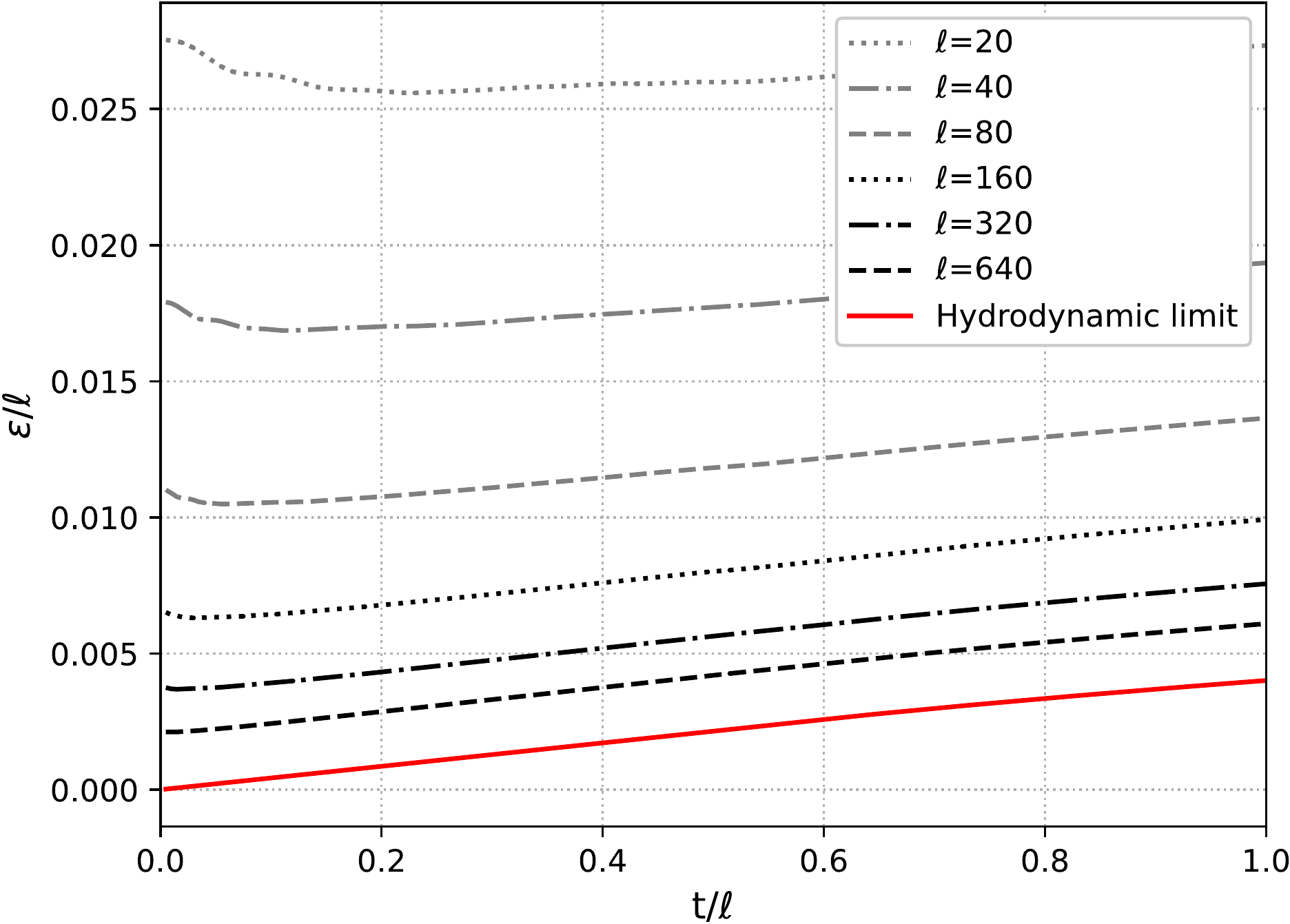}
	\caption{ Same as in Fig.~\ref{fig:conv_full} for $t/\ell\le1$ and larger 
		subsystem sizes $\ell$ up to $\ell=640$. 
	}
	\label{fig:conv_detail}
\end{figure}
%#############################################################
%
To confirm that, in Fig.~\ref{fig:conv_detail} we restrict ourselves to the region $t/\ell\le 1$ 
showing data up to $\ell=640$. As it is clear from the figure, the slope of the data with the largest 
$\ell=640$ is compatible with~\eqref{eq:main}. 

However, we should mention that corrections seem to decay in a slower manner as compared with 
the case of global dissipation~\cite{alba2022logarithmic}. 
To understand the origin of this fact, we now investigate the subleading terms in~\eqref{eq:main}. 
First, since the initial state is a Fermi sea, at $t=0$ the negativity exhibits 
logarithmic scaling~\cite{calabrese2012entanglement} with the size $\ell$. It is 
natural to conjecture that similar logarithmic scaling survives in the presence of loss. This suggests 
that 
\begin{equation}
\label{eq:e-asy}
{\cal E}={\cal E}_{\mathrm{asy}} + \alpha(k_F,\gamma^-)\ln(\ell)+{\mathcal O(1)}, 
\end{equation}
where ${\cal E}_\mathrm{asy}$ is given by~\eqref{eq:main}, and $\alpha(k_F,\gamma^-)$ is 
a constant, which in principle depends on $\gamma^-$ and $k_F$. Now, for $\gamma^-\to0$, the 
linear growth is absent, i.e., ${\mathcal E}_\mathrm{asy}\to0$, and~\cite{calabrese2012entanglement}  
\begin{equation}
\label{eq:central}
\alpha(k_F,0)\to \frac{c}{4}, 
\end{equation}
where $c=1$ is the central charge of the Conformal Field Theory~\cite{difrancesco1997conformal} 
(CFT) describing the long-wavelength properties of the Fermi sea. Notice that~\eqref{eq:central} 
does not depend on $k_F$. 
Similar logarithmic corrections as in~\eqref{eq:e-asy} 
should be expected for the von Neumann entropy~\cite{alba2021unbounded}. 

%
%#############################################################
\begin{figure}
	\centering
	\includegraphics[width=0.65 \linewidth]{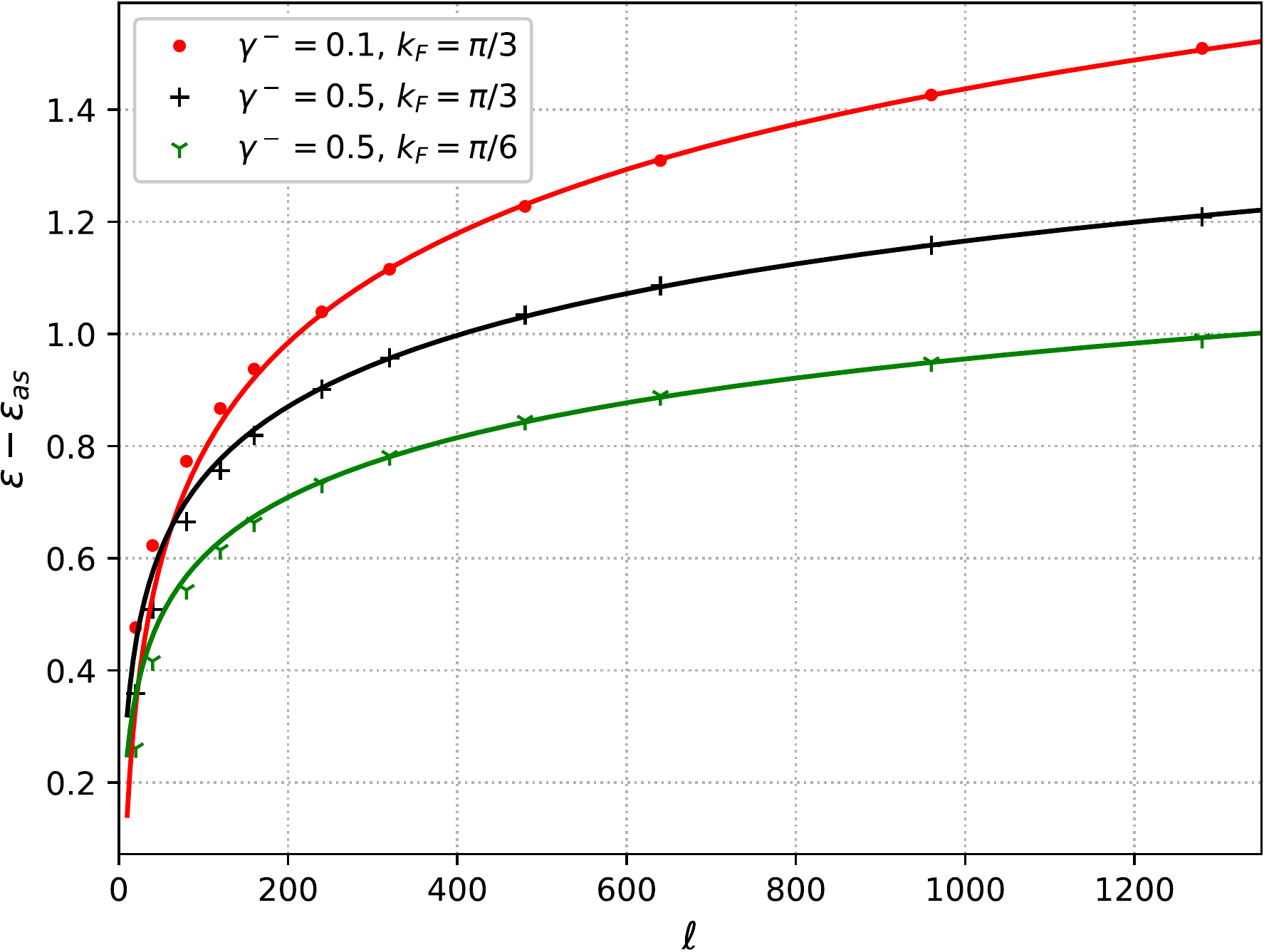}
	\caption{Logarithmic negativity ${\cal E}$ in a free fermion chain with localized loss:  
		Deviations from the hydrodynamic limit. We plot ${\cal E}-{\cal E}_\mathrm{as}$, with 
		${\cal E}_\mathrm{as}$ the analytic result in the hydrodynamic limit (cf.~\eqref{eq:main}), 
		versus $\ell$. Data are for fixed $t/\ell=0.5$. The symbols are data for different 
		loss rate $\gamma^-$ and Fermi momentum $k_F$. The continuous lines are fits to 
		${\cal E}-{\cal E}_\mathrm{as}=c\ln(\ell)+d$, with $c,d$ fitting parameters. 
	}
	\label{fig:subleading_lin}
\end{figure}
\begin{figure}
	\centering
	\includegraphics[width=0.65 \linewidth]{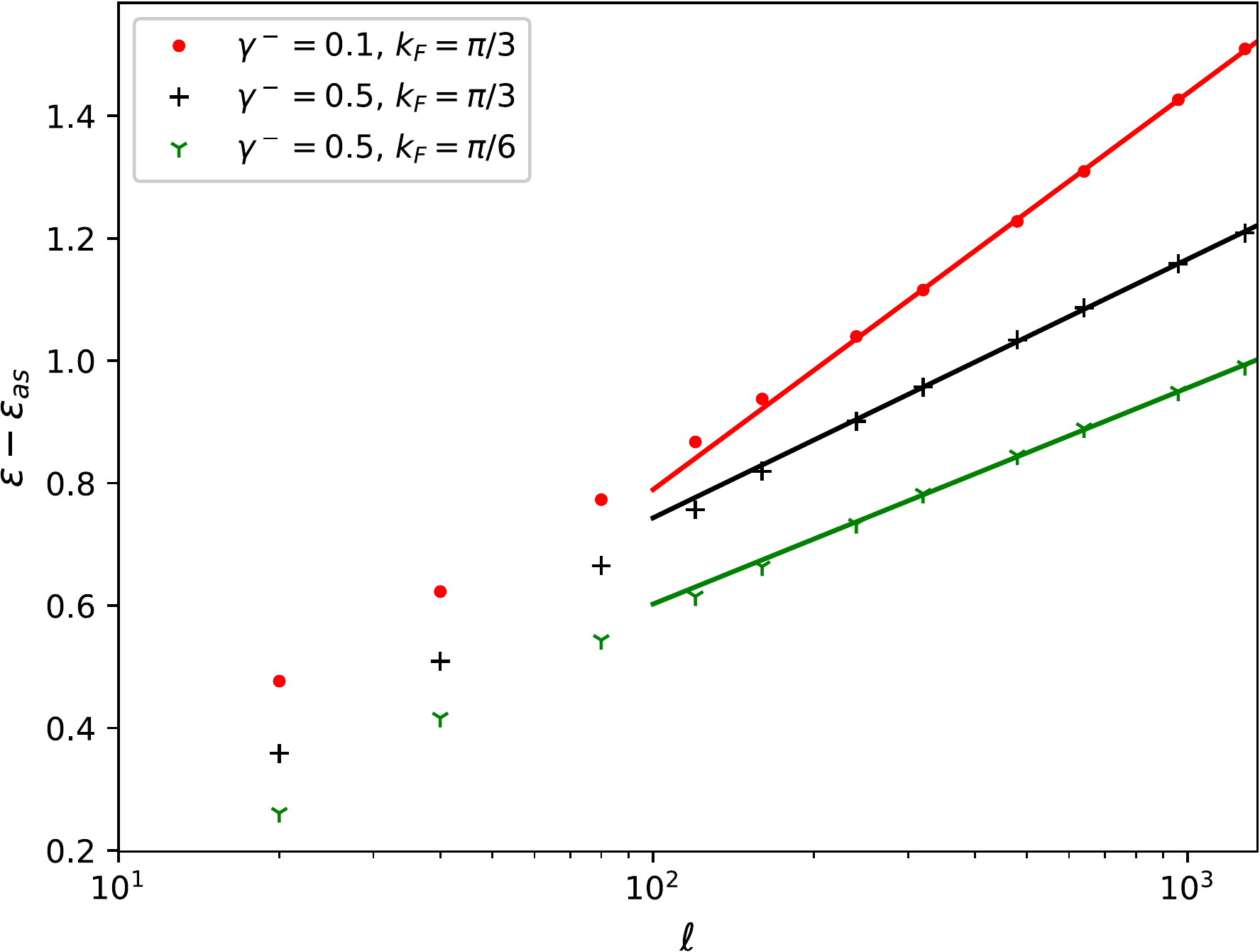}
	\caption{ Same as in Fig.~\ref{fig:subleading_lin} using a logarithmic scale on the 
		$x$-axis. Notice the dependence on the loss rate $\gamma^-$ of the prefactor of the 
		logarithmic growth. 
	}
	\label{fig:subleading_log}
\end{figure}

%#############################################################
%
To check the validity of~\eqref{eq:e-asy} we fix $t/\ell=0.5$ plotting the difference ${\cal E}-{\cal E}_\mathrm{asy}$ 
versus $\ell$. This is reported in Fig.~\ref{fig:subleading_lin} for several values of $k_F$ and $\gamma^-$. 
Clearly, ${\cal E}-{\cal E}_\mathrm{asy}$ exhibits a mild increase upon increasing $\ell$. 
The continuous lines are fits to the logarithmic increase ${\cal E}-{\cal E}_\mathrm{asy}=c\ln(\ell)+d$, 
with $c,d$ fitting parameters. The quality of the fits is satisfactory already for $\ell\gtrsim 200$, 
supporting the conjectured expression~\eqref{eq:e-asy}. 
Typical fitted values for $c,d$ are $c\simeq 0.2, d \simeq -0.1$ for $\gamma^-=0.5$, $k_F=\pi/3$. 
Interestingly, $\alpha$ exhibits a nontrivial dependence on 
$\gamma^-$. This is better illustrated in Fig.~\ref{fig:subleading_log} plotting the same data 
using a logarithmic scale on the $x$-axis. The different slopes that we 
observe for the data with 
fixed $k_F=\pi/3$ and $\gamma^-=0.5$, $\gamma^-=0.1$ suggest that $\alpha$ depends on $\gamma^-$. Notice 
also that the comparison between the data with $k_F=\pi/3$ and $k_F=\pi/6$ at fixed $\gamma^-=0.5$ 
suggests that $\alpha$ depends on $k_F$, although this could be a finite-size effect. 
Finally, one should remark that similar subleading 
logarithmic terms have been observed in the scaling of the von Neumann entropy in the presence of 
defects~\cite{fraenkel2022extensive}.
%
%#############################################################
%#############################################################
%

%##################################

%##################################
\section{Conclusion}
\label{sec:conclusions}

We investigated genuine entanglement propagation in a one-dimensional Fermi 
sea subject to both unitary and dissipative dynamics due to the presence of a localized loss. 
Our main result is formula~\eqref{eq:main}. Our formula shows that in the standard hydrodynamic 
limit of large intervals and long times, with their ratio fixed, the fermionic logarithmic 
negativity between two subsystems placed across the impurity grows linearly with time at short times, 
and it saturates to a volume-law scaling at asymptotically long times. The genuine entanglement 
production is traced back to the presence of maximally entangled pairs formed by the reflected 
and transmitted fermions scattered by the impurity. Although the derivation of~\eqref{eq:main} 
is quite involved, we provide a simple physical interpretation in terms of a two-level 
system for the two intervals. Our formula holds for the geometry with two adjacent intervals, 
although the generalization to more complicated geometries, such as the one with two disjoint 
intervals is straightforward. 

There are several promising avenues for future research. First, here we considered the paradigmatic 
setup with a uniform Fermi sea. It would be interesting to consider inhomogeneous initial states~\cite{alba2022noninteracting}, 
and more general localized dissipators, such gains/losses dissipators. An important question is 
whether the simple two-level system interpretation of the negativity discussed in section~\ref{sec:main-result} can be 
generalized. Clearly, it would be interesting 
to go beyond quadratic dissipative systems, for instance, by considering the effect of 
localized dephasing or incoherent hopping. The Lindblad equation for both these types of dissipation  
is integrable by using Bethe ansatz~\cite{eisler2011crossover,medvedyeva2016exact}. It would be important 
to investigate whether integrability persists for local dephasing and incoherent hopping. Our analysis revealed that the fermionic  negativity exhibits subleading logarithmic 
terms besides the leading hydrodynamic one, reflecting  that the initial state is critical. 
It would be interesting to characterize these corrections. This would allow to understand the interplay 
between dissipation and criticality. 

Finally, in our work we considered the dynamical regime in which 
$\ell,t\to \infty$ and $t/\ell$ is fixed. This is only one possibility and 
several different dynamical regimes can be studied~\cite{tarantelli2022out}. It would be 
interesting to characterize the entanglement dynamics in these regimes.

%##################################
\section*{References}
\bibliographystyle{iopart-num.bst}
\bibliography{bibliography}

\appendix

%##########################
\section{Hydrodynamic prediction for the negativity: Derivation of Eq.~\eqref{eq:main}} 
\label{sec:app-1}

Let us consider the case of the two symmetric intervals $A_1$ and $A_2$, with 
$A_1=[1,\ell]$ and $A_2=[-\ell,-1]$  as our subsystems 
(see Fig.~\ref{fig:cartoon}). The covariance matrix $G$ (cf.~\eqref{eq:Gxy}) has the structure:
\begin{equation}
	\label{eq:G-app}
	G=
	\begin{pmatrix}
		G^{11} & G^{12}\\
		G^{21}&G^{22}\\
	\end{pmatrix}=
	\begin{pmatrix}
	G_{x,y} & G_{x,-y}\\
	G_{-x,y}&G_{-x,-y}\\
\end{pmatrix},\quad x,y\in [1,\ell]. 
\end{equation}
In the same way and following section~\ref{sec:negativity}, we can use~\eqref{eq:G-app} in the definition of  
the  matrix $\Gamma= \mathds{1}-2G$ (cf.~\eqref{eq:gamma-a}) as 
\begin{equation}
	\label{eq:gamma-pm}
\Gamma^{\pm}_{x,y}:=2 \begin{pmatrix}
G_{x,y} & \mp iG_{x,-y}\\
\mp iG_{-x,y} & -G_{-x,-y}\\
\end{pmatrix} - \delta_{x,y}\sigma_z, \quad x,y\in [1,\ell], 
\end{equation}
where $\sigma_z$ is the Pauli matrix. 
We observe that $\Gamma^++\Gamma^-$ is block diagonal, i.e., 
\begin{equation}
	\label{eq:gp+gm}
\Gamma^+_{x,y} + \Gamma^-_{x,y}=4 \begin{pmatrix}
G_{x,y} & 0\\
0 & -G_{-x,-y}\\
\end{pmatrix} - 2\delta_{x,y}\sigma_z. 
\end{equation}
Using the definition of $G_{x,y}$ (cf.~\eqref{eq:Gxy}), we 
can now write $\Gamma_{x,y}^\pm$ as
\begin{equation}
	\label{eq:Gamma-int}
\Gamma^\pm_{x,y}= 2\int_{-k_F}^{k_F} \frac{dk}{2\pi}\begin{pmatrix}
S_{k,x}\\ \mp i S_{-k,x}\\
\end{pmatrix} \begin{pmatrix}
\bar{S}_{k,y} & \mp i \bar{S}_{-k,y}\\
\end{pmatrix}
-\delta_{x,y}\sigma_z
\end{equation}
where $S_{k,x}$ and $S_{k,y}$ are defined in~\eqref{eq:S}. 
We also have
\begin{equation}
	\label{eq:Gamma-defect-int}
\Gamma_{x,y}^++\Gamma_{x,y}^-=4\int_{-k_F}^{k_F} \frac{dk}{2\pi}\begin{pmatrix}
S_{k,x}\bar{S}_{k,y} & 0\\
0 &- S_{-k,x}\bar{S}_{-k,y}\\
\end{pmatrix}
-2\delta_{x,y}\sigma_z
\end{equation}
where we used the fact that $S_{k,-x}=S_{-k,x}$, which is apparent from~\eqref{eq:S}.

To compute the fermionic negativity, it is necessary to evaluate the moments of the 
correlators $M_p(\{n_j\})$ defined as 
\begin{equation}
	\label{eq:mom}
	M_p(\{n_j\})=\mathrm{Tr}\left[\prod_{j=1}^p
	(\Gamma^+ \Gamma^-)^{n_j}(\Gamma^+ + \Gamma^-)\right]. 
\end{equation}
Here $n_j$ with  $j=1,\dots,p$ is a positive integer. 
The matrices $\Gamma^\pm$ appearing in~\eqref{eq:mom} 
are restricted to subsystem $A$, i.e., they are $2\ell\times 2\ell$ 
matrices. Moreover, $M_p$ depends on time. Here we are interested 
in the hydrodynamic limit, which is defined as the limit $t,\ell\to\infty$, 
with the ratio $t/\ell$ fixed. Terms of the form~\eqref{eq:mom-trunc} are obtained from the 
series representation of the matrix $\Xi$ (cf.~\eqref{eq:xi-def}). 

%##########################
\subsection{Truncated moments} 
\label{sec:trunc}
To proceed, it is crucial to observe that $\Gamma^\pm$ 
(cf.~\eqref{eq:gamma-pm}) consists of the sum of two terms. The first one 
depends on $S_{k,x}$ (see~\eqref{eq:S} and~\eqref{eq:Gamma-int}), whereas the second one is 
$\delta_{x,y}\sigma_z$ and does not depend on the quasimomenta. 
For now, in multiplying the matrices $\Gamma^\pm$ 
in~\eqref{eq:mom}, we are going to neglect the term $\sigma_z$ 
in~\eqref{eq:gamma-pm}.  Let us define these ``truncated'' moments as 
$M_p^t$. They are defined as 
\begin{equation}
\label{eq:mom-trunc}
M^t_p(\{n_j\})=\mathrm{Tr}\left[\prod_{j=1}^p
(\Gamma^+ \Gamma^-)^{n_j}(\Gamma^+ + \Gamma^-)\right]^t, 
\end{equation}
where the superscript $t$ in the right hand side is to stress that we neglect the term $\delta_{x,y}\sigma^z$ in 
the definition of $\Gamma^\pm$ (cf.~\eqref{eq:gamma-pm}). 
The trace in~\eqref{eq:mom-trunc} is performed over both the spatial indices  as well as over the 
indices of the $2\times2$ block matrix. After performing 
the trace over the latter, a simple structure arises. 
Precisely, upon expanding the product in~\eqref{eq:mom-trunc} one obtains a  
string of operators $\Gamma^\pm$, which gives rise to 
strings of $S_{k,x}$. The strings of $S_{k,x}$ are  
constructed following three rules that we now discuss. 
Let us consider a generic string ${\cal S}$ of operators as 
\begin{equation}
\label{eq:string}
{\cal S}=(\Gamma^{+}\Gamma^-)^{n_1}(\Gamma^++\Gamma^-)(\Gamma^+\Gamma^-)^{n_2}\cdots 
(\Gamma^+\Gamma^-)^{n_p}(\Gamma^++\Gamma^-). 
\end{equation}
Let us recall that each $\Gamma^{\pm}$ contains an integral over one of 
the quasimomenta $k_j$ (cf.~\eqref{eq:gamma-pm}), 
with $k_1$ the quasimomentum associated with the leftmost $\Gamma^+$ in the string~\eqref{eq:string}, and $k_{2N_p+p}$ the one associated with 
the last term $(\Gamma^++\Gamma^-)$. 
The rules to expand~\eqref{eq:string} in terms of strings of $S_{k,x}$ (cf.~\eqref{eq:S}) are the 
following. 
\begin{itemize}
\item[$(i)$] There is an overall factor $2^{2N_p+2p}$, with $N_p=\sum_j n_j$, and 
a multidimensional integral over the quasimomenta $k_j$, with $j=1,2,\dots, 2N_p+p$ as 
\begin{equation}
\label{eq:r1}
M_p^t\to 2^{2N_p+2p}\int_{-k_F}^{k_F}\frac{d^{2N_p+p}k}{(2\pi)^{2N_p+p}} I(\{k_j\}), 
\end{equation}
where $I(\{k_j\})$ is determined from~\eqref{eq:string} by rules $(ii)$ and $(iii)$ below. 
\item[$(ii)$] Every couple of neighbouring operators 
$\Gamma^+_{x_1,x_2}(k_1) \Gamma^-_{x_2,x_3}(k_2)$ or 
$\Gamma^-_{x_1,x_2}(k_1) \Gamma^+_{x_2,x_3}(k_2)$ that one encounters by 
scanning site by site the string ${\cal S}$ in~\eqref{eq:string} from left to right contributes with a 
term $(\bar{S}_{k_1,x_2} S_{k_2,x_2} + \bar{S}_{-k_1,x_2} S_{-k_2,x_2} )$ to the integrand $I(\{k_j\})$ in~\eqref{eq:r1} (cf.~\eqref{eq:Gamma-int}).
\item[$(iii)$] Each sequence  of the type $\Gamma^-_{x_1,x_2} (k_1)(\Gamma^+_{x_2,x_3}(k_2) + \Gamma^-_{x_2,x_3}(k_2)) 
\Gamma^+_{x_3,x_4}(k_3)$ that one can identify in ${\cal S}$ contributes with a term $(\bar{S}_{k_1,x_2} S_{k_2,x_2}\bar{S}_{k_2,x_3} S_{k_3,x_3}  - \bar{S}_{-k_1,x_2} S_{-k_2,x_2}\bar{S}_{-k_2,x_3} 
S_{-k_3,x_3})$ in $I(\{k_j\})$ (see~\eqref{eq:Gamma-int} and~\eqref{eq:Gamma-defect-int}).
\end{itemize}
In rule $(ii)$ and $(iii)$ we sum over repeated spatial indices $x_j$. Notice also that 
the symbol $\Gamma^\pm(k_j)$ with the argument $k_j$ denotes the integrand in the definition 
of $\Gamma^\pm$ (cf.~\eqref{eq:Gamma-int}) and $k_j$ is the integration variable. 
Let us check the combined effect of rule $(ii)$ and $(iii)$ by considering the case with 
$p=2$ and $n_1=n_2=1$ in~\eqref{eq:mom-trunc}. The string ${\cal S}$ in~\eqref{eq:string} 
becomes 
\begin{multline}
\label{eq:M-2}
{\cal S}=\Gamma_{x_1,x_2}^+(k_1)\Gamma^-_{x_2,x_3}(k_2)(\Gamma_{x_3,x_4}^+(k_3)+
\Gamma_{x_3,x_4}^-(k_3))\\\times\Gamma^+_{x_4,x_5}(k_4)\Gamma_{x_5,x_6}^-(k_5)
(\Gamma^+_{x_6,x_1}(k_6)+\Gamma_{x_6,x_1}^-(k_6)). 
\end{multline}
The application of rule $(ii)$ and $(iii)$ gives 
\begin{multline}
\label{eq:last}
M_2^t\to
(\bar S_{k_1,x_2}S_{k_2,x_2}+\bar S_{-k_1,x_2}S_{-k_2,x_2})
(\bar{S}_{k_2,x_3} S_{k_3,x_3}\bar{S}_{k_3,x_4} S_{k_4,x_4}  - 
\bar{S}_{-k_2,x_3} S_{-k_3,x_3}\bar{S}_{-k_3,x_4} 
S_{-k_4,x_4})\\
(\bar S_{k_4,x_5}S_{k_5,x_5}+\bar S_{-k_4,x_5}S_{-k_5,x_5})
(\bar{S}_{k_5,x_6} S_{k_6,x_6}\bar{S}_{k_6,x_1} S_{k_1,x_1}  - 
\bar{S}_{-k_5,x_6} S_{-k_6,x_6}\bar{S}_{-k_6,x_1} 
S_{-k_1,x_1})
\end{multline}
Again, in~\eqref{eq:M-2} and~\eqref{eq:last} repeated spatial indices are 
summed over.  As it is clear from~\eqref{eq:last}, the 
effect of the operators $(\Gamma^++\Gamma^-)$ is to introduce the minus signs in 
the terms in the round brackets. This is due to the fact that upon expanding Eq.~\eqref{eq:M-2} there are 
terms with ``defects'', i.e., that contain the same operators on consecutive places. 
Before proceeding, it is useful to notice that 
if $p$ in~\eqref{eq:mom-trunc} is odd, the integration of the 
resulting product vanishes. Indeed,  the integration domain is invariant 
under the transformation $k_i \rightarrow -k_i \; \forall i$, 
while the integrand is odd under this transformation. This happens because  
rule $(iii)$ provides $p$ terms which are odd, while rule $(ii)$ give only  even terms. 
To perform the summation over the spatial indices, we use the identity
\begin{equation}
\label{eq:id}
\sum_{z=1}^\ell e^{izk}=\frac{\ell}{4}\int_{-1}^1d\xi \frac{k}{\sin(k/2)}e^{i(\ell\xi+\ell+1)k/2}. 
\end{equation}
It allows to rewrite the correlator ~\eqref{eq:mom-trunc} 
as an integral in $2 N_p+p$ $\xi$-variables, alongside an equal number of quasimomentum $k$-variables.
We obtain, for the generic case:
\begin{multline}
\label{eq:ugly}
M_p^t=
\left(\frac{\ell}{2}\right)^{2N_p+p} \int_{-k_F}^{k_F} \frac{d^{2N_p+p}k}{(2\pi)^{2N_p+p}} \int_{-1}^{1} d^{2N_p+p}\xi 
\\
\sum_{\mathrm{signs}}\,\,\sum_{\sigma_j,\tau_j=0,1}  
\prod_{i=1}^{p} \Big[\Big(\prod_{ j=2N_{i-1}+i}^{J_i=2N_i+i-1} 
e^{i\ell(\xi_j+1)([\pm k]_{\sigma_j}-[\pm k]_{\tau_{j-1}})/2} \tilde{r}^{\sigma_j}\tilde{r}^{\tau_{j-1}}\Big)\\
\times\underline{\Big\{ (\pm 1)   e^{i\ell(\xi_{J_i+1}+1)([\pm k]_{\sigma_{J_i+1}}-
		[\pm k]_{\tau_{J_i}})/2} e^{i\ell(\xi_{J_i+2}+1)([\pm k]_{\sigma_{J_i+2}}-[\pm k]_{
			\tau_{J_i+1}})/2} \tilde{r}^{\sigma_{J_i+1}}\tilde{r}^{\tau_{J_i}}\tilde{r}^{\sigma_{J_i+2}}
	\tilde{r}^{\tau_{J_i+1}} \Big\}}\Big],
\end{multline}
where we defined $N_i:=\sum_{j=1}^i n_j$ (cf.~\eqref{eq:mom-trunc}), and 
\begin{equation}
\label{eq:abs-def}
[k]_{\sigma_j}=\left\{
\begin{array}{cc}
k_j & \sigma_j=0\\
|k_j| & \sigma_j=1
\end{array}
\right.
\end{equation}
Moreover, we defined $\tilde r$ as 
\begin{equation}
\label{eq:rtilde}
\tilde r^{\sigma_j}:=\left(r(k_j)\Theta(-\ell(\xi_j+1)/2+|v_{k_j}|t) \right)^{\sigma_j}. 
\end{equation}
Here $r(k_j)$ is the reflection coefficient in~\eqref{eq:r-coeff}. 
Finally, in~\eqref{eq:ugly} we sum over the $\pm$ signs in the term within 
the square brackets. Importantly, all the signs in the underlined term in the curly brackets 
in~\eqref{eq:ugly} have to be equal.\\
Let us now discuss the structure  of~\eqref{eq:ugly}. The integration in $\xi_j$ 
originates from the application of~\eqref{eq:id} to treat the matrix multiplications 
and the trace in~\eqref{eq:mom-trunc}. Crucially, in applying~\eqref{eq:id} we used that in the hydrodynamic limit $\ell,t\to\infty$ with 
their ratio $t/\ell$ fixed, the stationary phase approximation for the integral in~\eqref{eq:ugly} 
gives as stationary point $k_j\to \pm k_1$ for each $j$. Moreover, we used that 
the integral in $dq$ in the definition of $S_{k,x}$ (cf.~\eqref{eq:S-1}) is dominated by $q\to0$. 
This implies that we can replace $k/\sin(k/2)\to2$ in~\eqref{eq:id}. We can also perform the 
integrations over $q$, which after using the identity 
\begin{equation}
\label{eq:id-1}
\int_{-\infty}^\infty\frac{dq}{2\pi i}\frac{e^{iqx}}{q-i0}=\Theta(x),  
\end{equation}
give the terms with $\tilde r$ (cf.~\eqref{eq:rtilde}) in~\eqref{eq:ugly}. 

The sum over the variables $\sigma_j,\tau_j$ in~\eqref{eq:ugly} 
takes into account that each $S_{k,x}$ (cf.~\eqref{eq:S-1}) contains a term that 
depends on $k_j$ and one that depends on $|k_j|$. The latter  is associated with a factor $\tilde{r}(k_j)$. The sum over the signs of $k_j$ reflects that 
for each term  $S_{k,x}$ in~\eqref{eq:mom-trunc} there is a term with $S_{-k,x}$ 
(see, for instance, Eq.~\eqref{eq:last} for the case with $p=2$). 
The signs in the second row of~\eqref{eq:ugly} can be chosen independently for each exponential in the product, as rule $(ii)$ yields two possible factors with opposite quasi-momentum, 
while in the underlined term in~\eqref{eq:ugly} all the signs have to be 
equal, following from rule $(iii)$. Indeed,  in rule $(iii)$ the terms $S_{k,x}$ with the minus sign 
 have also the two quasimomenta reversed. 

To illustrate the structure of~\eqref{eq:ugly} let us consider again the 
case with $p=2$ and $n_1=n_2=1$. Now, Eq.~\eqref{eq:ugly} becomes 
\begin{multline}
\label{eq:M2-expand}
M_2^t(1,1)=\\
\left(\frac{\ell}{2}\right)^6\sum_\mathrm{signs}\int_{-k_F}^{k_F}\frac{d^6k}{(2\pi)^6}
\int_{-1}^1d^6\xi \sum_{\sigma_j,\tau_j}e^{i\ell(\xi_2+1)([\pm k]_{\sigma_2}-[\pm k]_{\tau_1})+
	i\ell(\xi_5+1)([\pm k]_{\sigma_5}-[\pm k]_{\tau_4})}
\tilde r^{\sigma_2}\tilde r^{\tau_1}\tilde r^{\sigma_5}\tilde r^{\tau_4}\\
\left\{(\pm 1)e^{i\ell(\xi_3+1)([\pm k]_{\sigma_3}-[\pm k]_{\tau_2})+
	i\ell(\xi_4+1)([\pm k]_{\sigma_4}-[\pm k]_{\tau_3})}\tilde r^{\sigma_4}\tilde r^{\sigma_3}
\tilde r^{\tau_2}\tilde r^{\tau_3}\right\}\\
\left\{(\pm 1)e^{+i\ell(\xi_6+1)([\pm k]_{\sigma_6}-[\pm k]_{\tau_5})
	+i\ell(\xi_1+1)([\pm k]_{\sigma_1}-[\pm k]_{\tau_6})}
\tilde r^{\sigma_6}\tilde r^{\sigma_1}
\tilde r^{\tau_5}\tilde r^{\tau_6}
\right\}
\end{multline}
Let us now perform the stationary phase analysis in the limit $\ell,t\to\infty$ with 
their ratio $t/\ell$ fixed. 
Here we perform the stationary phase approximation with respect to both variables 
$k_j$ and $\xi_j$. 
Before proceeding, let us restrict ourselves to the case in which the variables $\tau_j,\sigma_j$ 
are not all zero. The configuration with $\sigma_j=\tau_j=0,\,\forall j$ will be discussed at the 
end. 
First, we notice that all the quasimomenta appear twice in the phase factors in~\eqref{eq:ugly}, as it 
is also clear from~\eqref{eq:M2-expand} (notice the presence of the terms $[\pm k]_{\sigma_j}$ and $[\pm k]_{\tau_j}$). 
Let us define the two indices $(\sigma_j,\tau_j)$ as \emph{paired} if $\sigma_j=\tau_j=1$, and 
\emph{unpaired} otherwise. 
Moreover, we observe that we can treat the terms arising from the operators $\cdots(\Gamma^++\Gamma^-)\cdots$ 
and $\cdots(\Gamma^+\Gamma^-)^{n_j}\cdots$ separately. 
Let us start discussing the first ones, which corresponds to the underlined terms in the curly brackets in~\eqref{eq:ugly}. 
Now, the signs $\pm$ of two occurrences of  $k_j$ have to be the same because 
both occurrences of $k_j$ are in the 
same underlined block. This is also clear in~\eqref{eq:M2-expand} for $p=2$. The two occurrences of the 
quasimomenta $k_3$ and $k_6$ appear with the same sign. 

It is straightforward to check that if one has $\sigma_j=0$ or $\tau_j=0$, i.e., if 
the $(\sigma_j,\tau_j)$ are not paired, then the stationarity condition 
with respect to $k_j$ fixes the sign of $k_j$. Specifically, one has that 
$k_j>0$ if the chosen sign in the curly bracket to which $k_j$ belongs is plus, 
and minus otherwise. This follows from the requirement that the solution of the stationarity  
condition for $\xi_i$ belongs to the integration domain $[-1,1]$ $\forall i$. 
On the other hand, if the indices $(\sigma_j,\tau_j)$ are paired, 
then the sign of $k_j$ is not fixed, i.e., $k_j\in[-k_F,k_F]$. Since the integrand is 
symmetric under exchange $k_j\to-k_j$, the difference between paired and unpaired indices 
is a factor two. 
Now, the calculation of the contribution of the terms in the curly brackets in~\eqref{eq:ugly} 
involves some elementary combinatorics, and it is the same as 
in Ref.~\cite{alba2021unbounded}. For now, we also omit the step function in~\eqref{eq:rtilde}.  
In summary, the result is 
\begin{equation}
\label{eq:res-part}
\cdots(\Gamma^++\Gamma^-)\cdots\to (1+2r+2r^2)^p=(1-|a(k)|^2)^p
\end{equation}
In~\eqref{eq:res-part} $|a(k)|^2$ is the absorption coefficient~\eqref{eq:ak}, and we are using that in the stationary phase approximation 
$k_j=\pm k$  for any $j$ and that $|a(k)|^2=|a(-k)|^2$.

Let us now discuss the contribution of the terms $\cdots(\Gamma^+\Gamma^-)^{n_j}\cdots$ in~\eqref{eq:string}.  
Now the pairs of momenta appearing in the generic block $(\Gamma^+\Gamma^-)^{n_j}$ 
are not forced to have the same sign. Again, this is clear by considering the 
toy example with $p=2$ and $n_1=n_2=1$ in~\eqref{eq:M2-expand}. 
In principle, one has to sum over the plus and minus signs in $[\pm k]_{\sigma_j}$ and $[\pm k]_{\tau_j}$.  
The result depends on the choice of sign of the quasimomenta at the edges of the block. Let us 
first consider the case in which the quasimomenta at the edges have the same sign. 
Let us also start from the case in which all the $\sigma_j$ and $\tau_j$ associated with 
quasimomenta in the block are zero. Now, it 
is easy to convince oneself that stationarity with respect to $k_j$ 
implies that there is only one allowed sign configuration, which  gives result $1$. 
It also not difficult to check that if there is at least a 
$\sigma_j$ or a $\tau_j$ equal to $1$ there is a factor $2^{i-1}$, 
where $i$ is the the number of indices $\sigma_j,\tau_j$ that are equal to $1$. 
By summing over $\sigma_j,\tau_j$, one obtains $Z_{2n_j}$, where  
\begin{equation}
\label{eq:Z_nj}
Z_{n}:= \sum_{i=1}^{2n} \binom{2n}{i} 2^{i-1} r^i= \frac{(1+2r)^{2n}-1}{2}=\frac{(1-2|a|^2)^{n}-1}{2}
\end{equation}
Thus, the total result is $1+Z_{2n_j}$. Instead, if the signs of the quasimomenta at the edges 
are different, the result is $ Z_{2n_j}$, since there is no stationary phase contribution from the case of all the $\sigma_j$, $\tau_j$ in the block equal to zero. 

Finally, we have to glue together the contributions of the different blocks 
$\cdots(\Gamma^+\Gamma^-)^{n_j}$ with different  $j$. 
It is easy to show that the final result is
\begin{equation}
\label{eq:res-part-1}
\prod_{j=1}^p\left(\cdots (\Gamma^+\Gamma^-)^{n_j}
\cdots\right)	\to\mathrm{Tr}\left[\prod_{j=1}^{p} \begin{pmatrix}
1 + Z_{2n_j} & - Z_{2n_j}\\
Z_{2n_j} & -1 - Z_{2n_j}\\
\end{pmatrix} \right]
\end{equation}
where $Z_{2n_j}$ is defined in~\eqref{eq:Z_nj}.
The minus sign in the second 
column is the minus sign that appears in the  terms with reversed quasimomenta in rule $(iii)$. 
Notice that if $n_j=0$, $Z_{2n_j}=0$ and there are no configurations 
with different sign at the endpoints, as it should be: 
$n_j=0$ implies the presence of two consecutive 
$(\Gamma^+ + \Gamma^-)$ factors. 
Putting all together, we obtain the final result for $M_p^t$ with even $p$  as 
\begin{multline}
\label{eq:mt-final}
M_p^t=
2^{2(N_p+p)-1} 
\int_{-k_F}^{k_F} \frac{dk}{2\pi} 
\Big\{4\ell+\min(|v_k|t,\ell)\Big((1-|a|^2)^p \mathrm{Tr}\Big[\prod_{i=1}^{p} \begin{pmatrix}
1 + Z_{2n_i} & - Z_{2n_i}\\
Z_{2n_i} & -1 - Z_{2n_i}\\
\end{pmatrix}\Big]-2\Big)\Big\}. 
\end{multline}
The term $\min(|v_k|t,\ell)$ comes from the final integration of the 
step function in~\eqref{eq:rtilde} that one has to perform every time 
there is at least a $\sigma_j$ or a $\tau_j$ equal to $1$, in 
the same way as explained in~\cite{alba2021unbounded}. 
The two configurations we have for $\sigma_j=\tau_j=0 \quad \forall j$ 
(the signs all equal to plus or to minus) must be subtracted 
from the combinatorial factor and treated separately, 
yielding the term $4\ell$ in~\eqref{eq:mt-final}.

%##########################
\subsection{Full moments} 
\label{sec:full-moments}

We now consider the full moments $M_p$, taking into account the terms $\delta_{x,y}\sigma_z$ 
in the definition of $G_{x,y}$ (cf.~\eqref{eq:gpm}).  This means that we are allowed to replace any 
occurrence of $G_{x,y}$ with $-\delta_{x,y}\sigma_z$ in each $\Gamma^\pm$ 
correlator. It is easy to check that by substituting a term 
$-\delta_{x,y}\sigma_z$ in a block  $(\Gamma^+\Gamma^-)^{n_j}$ the structure 
of the calculation remains the same as in the previous section. 
The result differs from~\eqref{eq:res-part-1} by a factor $-1/2$ and by 
the substitution $Z_{2n_j}\rightarrow Z_{2n_j -1}$. In a similar way, 
the replacement of a $G_{x,y}$ in a  term like $(\Gamma^++\Gamma^-)$ gives  
an overall $-1/2$ and the substitution 
$(1-|a^2|)^p\rightarrow (1-|a|^2)^{p-1}$ in~\eqref{eq:res-part}. 
So in the end Eq.~\eqref{eq:mt-final} becomes 
\begin{multline}
%\begin{gathered}
\label{eq:Mp}
M_p=\mathrm{Tr} \prod_{j=1}^p\left[(\Gamma^+ \Gamma^-)^{n_j}(\Gamma^+ + \Gamma^-)\right]
=\sum_{i=0}^{p}\binom{p}{i}  4^{p-i} (-2)^i \prod_{l=1}^p\sum_{j_l=0}^{2n_l} \binom{2n_l}{j_l} 2^{2n_l-j_l} (-1)^{j_l}\\
\times\left(\frac{\ell}{2}\right)\int_{-k_F}^{k_F} \frac{dk}{2\pi} \Big\{4+\min(|v_k|t/\ell,1)\Big((1-|a|^2)^{p-i} 
\mathrm{Tr}\Big[\prod_{h=1}^{p} \begin{pmatrix}
1 + Z_{2n_h-j_h} & - Z_{2n_h-j_h}\\
Z_{2n_h-j_h} & -1 - Z_{2n_h-j_h}\\
\end{pmatrix}\Big]-2 \Big)\Big\}.
\end{multline}
Here the first sum takes into account the different ways of inserting 
the term $\sigma_z$ in strings of the type $\cdots(\Gamma^++\Gamma^-)\cdots$. 
The second sum accounts for insertions in blocks of the 
type $\cdots(\Gamma^+\Gamma^-)^{n_j}\cdots$. 
This holds for $p$ even, while for odd $p$ every term in the sum 
is still zero, for the same symmetry reasons as above.
Before proceeding, we notice that in~\eqref{eq:Mp} the term with $i=p$ and 
$j_m=2n_m$ has to be treated separately. It corresponds to replacing all the 
occurrences of the operators $\Gamma^\pm$ with 
$\Gamma^\pm_{x,y}\sim \delta_{x,y}\sigma_z$. This term 
gives $2\ell$. Now, after performing the sums in~\eqref{eq:Mp}, we obtain: 
\begin{multline}
\label{eq:Mp1}
M_p=2^p \int_{-k_F}^{k_F} \frac{dk}{2\pi} \left\{\frac{2\pi\ell}{k_F}+\min(|v_k|t,\ell)
\Big(\frac{1}{2}
(1-2|a|^2)^p \mathrm{Tr}\Big[\prod_{h=1}^{p} \begin{pmatrix}
1 + \widetilde{Z}_{2n_h} & - \widetilde{Z}_{2n_h}\\
\widetilde{Z}_{2n_h} & -1 - \widetilde{Z}_{2n_h}\\
\end{pmatrix}\Big]-1 \Big)\right\} ,
%\end{gathered}
\end{multline}
where we have defined: 
\begin{equation}
	\widetilde{Z}_{n}:=\frac{(1-4|a|^2)^{n}-1}{2}.
\end{equation} 
The first term in the square brackets in~\eqref{eq:Mp1} corresponds to $i=p$ and $j_m=2n_m$. 

Eq.~\eqref{eq:Mp1} provides the building block to compute terms $\widetilde M_p$ of the form
\begin{equation}
\label{eq:tildeMp}
\widetilde M_p:=\mathrm{Tr}\left[ \left( P^{-1}(\Gamma^++\Gamma^-)\right)^p \right],  
\end{equation}
where $P=\mathds{1}+\Gamma^+\Gamma^-$. These terms are needed to compute the 
negativity~\eqref{eq:neg-def}. 
The strategy to compute~\eqref{eq:tildeMp} is to use the trivial identity 
\begin{equation}
P^{-1}=\sum_{n=0}^{\infty} (-1)^n (\Gamma^+ \Gamma^-)^n. 
\end{equation}
We can rewrite~\eqref{eq:tildeMp} as 
\begin{equation}
\label{eq:tildeMp1}
\widetilde M_p=\mathrm{Tr}\left[ \prod_{j=1}^p \left(\sum_{n_j=0}^\infty (-1)^{n_j}(\Gamma^+\Gamma^-)^{n_j}(\Gamma^++\Gamma^-)\right) \right]. 
\end{equation}
Now we can apply the result~\eqref{eq:Mp1} to each term in~\eqref{eq:tildeMp1}. 
By using the regularisation $\sum_{n=0}^{\infty} (-1)^n=\frac{1}{2}$, we obtain:
\begin{multline}
\label{eq:tildeMp2}
%\begin{gathered}
\widetilde M_p
=\int_{-k_F}^{k_F} \frac{dk}{2\pi} \left\{ \frac{2\pi\ell}{k_F} -\min(|v_k|t,\ell) 
+2^{p-1}\min(|v_k|t,\ell)(1-2|a|^2)^p \mathrm{Tr}\left[\begin{pmatrix}
1/2 + b & - b\\
b & -1/2 - b\\
\end{pmatrix}^p \right] \right\}=
\\
=\int_{-k_F}^{k_F} \frac{dk}{2\pi} \left[\frac{2\pi\ell}{k_F} -\min(|v_k|t,\ell) +
\min(|v_k|t,\ell)\left(\frac{1-2|a|^2}{\sqrt{(1-2|a|^2)^2+4|a|^4}}\right)^p \right],
%\end{gathered}
\end{multline}
where in the first row we defined 
\begin{equation}
b:= \sum_{n=0}^{\infty} (-1)^n \widetilde Z_{2n} = 
\frac{1}{2(1+(1-4|a|^2)^2)} -\frac{1}{4},
\end{equation}
Here we should stress that Eq.~\eqref{eq:tildeMp2} holds for even $p$, whereas for odd $p$ one has 
that $\widetilde M_p$ vanishes.

%##########################
\subsection{Fermionic negativity} 
\label{sec:ferm-neg}

We now have all the ingredients to compute the fermionic negativity. 
First, we notice that by using the results of section~\ref{sec:full-moments}, 
we can compute $\mathrm{Tr}[\mathcal{F}(P^{-1}(\Gamma^++\Gamma^-))]$ 
for a generic function ${\mathcal F}(z)$ admitting a Taylor 
expansion around $z=0$.
After expanding ${\mathcal F}(z)$ around $z=0$ and 
using~\eqref{eq:tildeMp2} for each term in the expansion, we obtain that 
\begin{multline}
%\begin{gathered}
\mathrm{Tr}\left[{\mathcal F}(P^{-1}(\Gamma^+ + \Gamma^-))\right]
=\frac{1}{2}\int_{-k_F}^{k_F} \frac{dk}{2\pi} \Big\{\left(\frac{2\pi\ell}{k_F} -\min(|v_k|t,\ell)\right) 
({\mathcal F}(1)+{\mathcal F}(-1))\\ 
+\min(|v_k|t,\ell)\Big[{\mathcal F}\Big(\frac{1-2|a|^2}{\sqrt{(1-2|a|^2)^2+4|a|^4}}\Big) 
+ {\mathcal F}\Big(-\frac{1-2|a|^2}{\sqrt{(1-2|a|^2)^2+4|a|^4}}\Big)\Big]\Big\}
%\end{gathered}
\end{multline}
The presence of the terms ${\mathcal F}(z)+{\mathcal F}(-z)$ 
follows from the fact that $\widetilde{M}_p=0$ for odd $p$. 
Clearly, by choosing ${\mathcal F }(z)=z^p$, one recovers~\eqref{eq:tildeMp2}. 
To compute  the negativity, we have to use the function (see the first term 
in~\eqref{eq:neg-def}) 
\begin{equation}
\label{eq:F-neg}
{\mathcal F}(z)=\ln\Big[\Big(\frac{1+x}{2}\Big)^{1/2}+\Big(\frac{1-x}{2}\Big)^{1/2}\Big]. 
\end{equation}
The second term in the definition of the negativity (see Eq.~\eqref{eq:neg-def}) is 
obtained from the results of Ref.~\cite{alba2021unbounded}. 
Putting everything together, we obtain that 
\begin{equation}
\label{eq:neg-final}
\mathcal{E}=
\frac{1}{2}\int_{-k_F}^{k_F} \frac{dk}{2\pi} 
\min(|v_k|t,\ell) \ln\left(2|a|^2+\sqrt{(1-2|a|^2)^2+4|a|^4}\right). 
\end{equation}
Similar to the von Neumann entropy~\cite{alba2021unbounded}, the negativity depends only on the 
absorption coefficient $|a(k)|^2$ of the lossy site.

%##########################
\section{The case of  two unentangled intervals}
\label{sec:app-2}

\begin{figure}
	\begin{center}
		\includegraphics[width=0.5\linewidth]{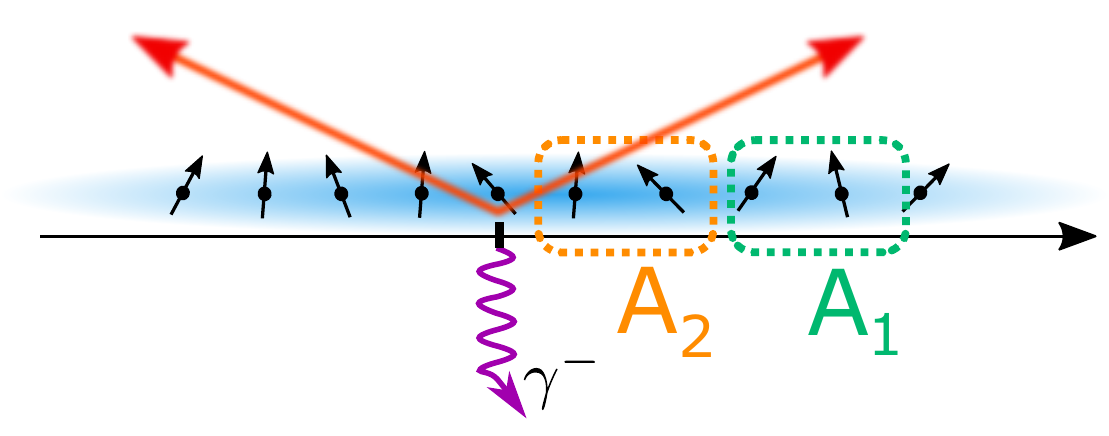}
		\caption{ Two adjacent equal-length intervals $A_1$ and $A_2$ of size $\ell$ placed 
			on the same side of the dissipative impurity. An entangled pair 
			formed by the reflected and transmitted fermions scattered by 
			the impurity is shown. As the two entangled particles are never 
			shared between the two intervals the negativity between them 
			is zero in the hydrodynamic limit. 
		}
		\label{fig:cartoon-app}
	\end{center}
\end{figure}

In this section we consider the tripartition in Fig.~\ref{fig:cartoon-app}. In contrast 
with Fig.~\ref{fig:cartoon} the two subsystems $A_1$ and $A_2$ are on the same side of the 
impurity. 
To be specific, let us take the two consecutive intervals 
$A_1=[1,\ell]$ and $A_2=[\ell+1,2\ell]$ as our subsystems. 
According to the scenario discussed in~\ref{sec:main-result} 
(see also Fig.~\ref{fig:cartoon} (b)), the entanglement between two intervals is due to 
the pairs of entangled quasiparticles produced at the impurity, and shared between 
the intervals. As it is clear from Fig.~\ref{fig:cartoon-app} this implies that 
the two intervals $A_1$ and $A_2$ cannot be entangled because the pairs of 
quasiparticles produced at the impurity site are never shared between them. 

It is instructive to prove this result using the approach of~\ref{sec:app-1}. 
%
%#############################################################
%#############################################################
%
The fermionic correlation matrix $G$ for $A_1\cup A_2$ is given as 
\begin{equation}
\label{eq:gxy-p}
G=\begin{pmatrix}
G^{11} & G^{12}\\
G^{21}&G^{22}\\
\end{pmatrix}=
\begin{pmatrix}
G_{x,y} & G_{x,y+\ell}\\
G_{x+\ell,y}&G_{x+\ell,y+\ell}\\
\end{pmatrix}, \quad x,y \in [1,\ell].
\end{equation}
Notice that~\eqref{eq:gxy-p} is different from~\eqref{eq:G-app}.
As in the previous section, we can write 
\begin{equation}
\label{eq:Gammapm}
\Gamma^{\pm}_{x,y}:=2 \begin{pmatrix}
G_{x,y} & \mp iG_{x,y+\ell}\\
\mp iG_{x+\ell,y} & -G_{x+\ell,y+\ell}\\
\end{pmatrix} - \delta_{x,y}\sigma_z, \quad x,y \in [1,\ell].
\end{equation} 
We now proceed as in section~\ref{sec:app-1}. We consider the moments 
\begin{equation}
\label{eq:mt}
\widehat M_p(\{n_j\}):=\mathrm{Tr}\left[\prod_{j=1}^{p}((\Gamma^+\Gamma^-)^{n_j}\Gamma^\pm)\right],  
\end{equation}
and their truncated version $\widehat M_p^t$ (cf.~\eqref{eq:mom-trunc}), 
where the superscript $t$ is to stress that we neglect the term 
$\sigma_z$ in~\eqref{eq:Gammapm}. 
Notice that, in contrast with the similar definition~\eqref{eq:mom-trunc}, here we 
do not sum over the $\pm$ in $\Gamma^\pm$ (notice the term $\Gamma^++\Gamma^-$ in~\eqref{eq:mom-trunc}).  
Using the definition of $G_{x,y}$ in~\eqref{eq:G-app}, we have 
\begin{equation}
\label{eq:gpmapp}
\Gamma^\pm_{x,y}= 2\int_{-k_F}^{k_F} \frac{dk}{2\pi} S^\pm_{k,x} \left({S^\pm_{k,y}}\right)^{\dagger},
\end{equation}
where we have defined $S^\pm_{k,x}$ as 
\begin{equation}
\label{eq:spmapp}
S^\pm_{k,x} := \begin{pmatrix}
S_{k,x}\\\mp i S_{k,x+\ell}
\end{pmatrix}=e^{ikx} \begin{pmatrix}
1\\ \mp i e^{ik\ell}
\end{pmatrix} + r(k) e^{i|k|x}\int_{-\infty}^{\infty}\frac{dq}{2\pi i} \frac{e^{i(|v_k|t - x)q}}{q-i0^+} \begin{pmatrix}
1\\ \mp i e^{i(|k|-q)\ell}
\end{pmatrix}, 
\end{equation}
where $r(k)$ is the reflection coefficient (cf.~\eqref{eq:r-coeff}), and the integral 
over $q$ yields the proper step functions~\eqref{eq:id-1}. 
After substituting~\eqref{eq:gpmapp} and~\eqref{eq:spmapp} in~\eqref{eq:mt} 
for each couple of $\Gamma$s in~\eqref{eq:mt} we have to evaluate sums of the type 
\begin{equation}
\label{eq:sum1-2}
\Sigma_1:=\sum_{x=1}^{\ell}  \left({S^\pm_{k,x}}\right)^{\dagger} \;S^\mp_{k',x},
\quad \Sigma_2:=\sum_{x=1}^{\ell}  \left({S^\pm_{k,x}}\right)^{\dagger} \;S^\pm_{k',x}. 
\end{equation}
These sums can be rewritten by using the identity~\eqref{eq:id}. 
We obtain terms of the type 
\begin{align}
\label{eq:f1}
& F^{uu}_{k_{i},k_{j}}=\frac{\ell}{4}\int_{-1}^{1} d\xi w(k_i - k_j) e^{i\ell(\xi +1)(k_i-k_j)/2} (1\pm e^{i\ell(k_i-k_j)}),\\
\label{eq:f2}
& F^{ud}_{k_{i},k_{j}}=-\frac{\ell}{4}\int_{-1}^1 d\xi \int \frac{dq}{2\pi i} w(k_i - |k_j| +q) e^{i\ell(\xi +1)(k_i-|k_j|+q)/2}\;
{A}^*_{k_j,q}  (1\pm e^{i\ell(k_i-|k_j|+q)}),\\
\label{eq:f3}
& F^{du}_{k_{i},k_{j}}=\frac{\ell}{4}\int_{-1}^1 d\xi \int \frac{dq}{2\pi i} w(|k_i| - k_j -q) 
e^{i\ell(\xi +1)(|k_i|-k_j-q)/2}\;
A_{k_i,q}  (1\pm e^{i\ell(|k_i|-k_j-q)}), 
\end{align}
and 
\begin{multline}
\label{eq:f4}
F^{dd}_{k_{i},k_{j}}=-\frac{\ell}{4}\int_{-1}^1 d\xi \int \frac{dq}{2\pi i} 
\int \frac{dq'}{2\pi i} \\w(|k_i| -| k_j| -q+q') e^{i\ell
	(\xi +1)(|k_i|-|k_j|-q+q')/2}\; A_{k_i,q} {A}^*_{k_j,q'} (1\pm e^{i\ell(|k_i|-|k_j|-q+q')}),
\end{multline}
where $w(k)=k/\sin(k/2)$, and we defined 
$A_{k,q}:=r(k)e^{it|v_k|q}/(q-i0^+)$ (the star denoting 
complex conjugation). The superscripts $u$ and $d$, for example, in $F^{ud}_{k_i,k_j}$ 
mean that in multiplying $(S^\alpha)^\dagger S^\beta$ (cf.~\eqref{eq:sum1-2}), with 
$\alpha,\beta=\pm$ we consider the first term in~\eqref{eq:spmapp} for $S^\alpha$ 
and the second one for $S^\beta$. The $u$ refers to the fact that the 
first term in~\eqref{eq:spmapp} 
is present also in the unitary case, whereas $d$ stands for ``dissipative'' because the 
second term in~\eqref{eq:spmapp} is due to the dissipation. 

The $\pm$ signs in~\eqref{eq:f1}\eqref{eq:f2}\eqref{eq:f3}\eqref{eq:f4} 
are fixed as follows: If we multiply $S^+S^-$ or $S^-S^+$ they are all $+$,
whereas if we multiply 
$S^+S^+$ or $S^-S^-$, we have all $-$ signs. 
The formulas~\eqref{eq:f1}\eqref{eq:f2}\eqref{eq:f3}\eqref{eq:f4} can be simplified in 
the hydrodynamic limit $t,\ell\to\infty$ with their ratio $t/\ell$ fixed. In this limit 
the integrals in~\eqref{eq:f1}-\eqref{eq:f4} are dominated by the points with $q,q'\to0$ and 
$k_j\to \pm \bar k$ for all $j$. This means that we can replace $w(k)\to 2$ and perform the integration 
over $q$. We obtain 
\begin{equation}
F^{uu}_{k_{i},k_{j}}=\frac{\ell}{2}
\int d\xi \; e^{i\ell(\xi +1)(k_i-k_j)/2} \pm e^{i\ell(\xi +3)(k_i-k_j)/2},
\end{equation}
\begin{multline}
F^{ud}_{k_{i},k_{j}}=\frac{\ell}{2}\int d\xi \; e^{i\ell(\xi +1)(k_i-|k_j|)/2}r(k_j)\Theta(-\ell(\xi+1)/2+|v_{k_j}|t)\\
\pm e^{i\ell(\xi +3)(k_i-|k_j|)/2}r(k_j)\Theta(-\ell(\xi+3)/2+|v_{k_j}|t),
\end{multline}
\begin{multline}
F^{du}_{k_{i},k_{j}}=\frac{\ell}{2}\int d\xi \; e^{i\ell(\xi +1)(|k_i|-k_j)/2}r(k_i)\Theta(-\ell(\xi+1)/2+|v_{k_i}|t)\\
\pm e^{i\ell(\xi +3)(|k_i|-k_j)/2}r(k_i)\Theta(-\ell(\xi+3)/2+|v_{k_i}|t),
\end{multline}
and
\begin{multline}
F^{dd}_{k_{i},k_{j}}=\frac{\ell}{2}\int d\xi \;e^{i\ell(\xi +1)(|k_i|-|k_j|)/2}r(k_i)r(k_j)\Theta(-\ell(\xi+1)/2+|v_{k_i}|t) 
\Theta(-\ell(\xi+1)/2+|v_{k_j}|t) \\
\pm e^{i\ell(\xi +3)(|k_i|-|k_j|)/2}r(k_i)r(k_j)\Theta(-\ell(\xi+3)/2+|v_{k_i}|t) \Theta(-\ell(\xi+3)/2+|v_{k_j}|t).
\end{multline}
The two sums $\Sigma_1$ and $\Sigma_2$ (cf.~\eqref{eq:sum1-2}) contain a term of the form $\sum_{\alpha,\beta} F^{\alpha\beta}_{k_i,k_j}$.

To proceed, let us observe that 
in calculating  the truncated moments $\widehat M_p^t$ (cf.~\eqref{eq:mt}), 
we have a multidimensional integral in the 
variables $k_j$ and $\xi_j$ with $j=1,\dots,2N_p + p$. 
We can now treat these integrals using the stationary phase 
approximation~\cite{wong}. The integrand contains products of the form: 
\begin{equation}
\label{eq:prodapp}
\prod_{i=1}^{2N_p+p} e^{i\ell
	\eta_i(k_{\sigma_i}-k_{\tau_{i-1}})/2} \; \tilde{r}^{\sigma_i}(\xi_i,k_{\sigma_i})\; \tilde{r}^{\tau_{i-1}}(\xi_i,k_{\tau_{i-1}}),
\end{equation}
where $\eta_i$ can be $\xi_i + 1$ or $\xi_i +3$, $k_{\sigma_i}$, $k_{\tau_i}$ are defined 
in~\eqref{eq:abs-def}, $\tilde{r}^{\sigma_i}$ are defined in~\eqref{eq:rtilde},  and 
$N_p:=\sum_{j=1}^p n_j$. 
Now, it is easy to verify that  the stationary point of~\eqref{eq:prodapp} 
is in the integration domain  
\textit{only} if all the $\eta_i$ are the same. 
For instance, if we have $\eta_i=\xi_i +1$ and $\eta_{i+1}=\xi_{i+1}+3$, in~\eqref{eq:prodapp} we have a 
term $((\xi_i +1)k_{\sigma_i}+(\xi_{i+1}+3) k_{\tau_i})/2$ in the exponent.  
Thus, the stationary point must satisfy the condition $\xi_i +1=\pm(\xi_{i+1}+3)$, which has no 
solution because $\xi_i$ are in $[-1,1]$.
This implies that 
\begin{multline}
\label{eq:Mtpapp}
\widehat M_p^t=\ell^{2N_p+p}\!\! \int_{-k_F}^{k_F}\!\! \frac{d^{2N_p+p} k}{(2\pi)^{2N_p+p}}\int\!\!\! d^{2N_p+p} \xi 
\sum_{\sigma_i,\tau_i} \prod_{i=1}^{2N_p+p} e^{i\ell(\xi_i+1)(k_{\sigma_i}-k_{\tau_{i-1}})/2} \; 
\tilde{r}^{\sigma_i}(\xi_i,k_{\sigma_i})\; \tilde{r}^{\tau_{i-1}}(\xi_i,k_{\tau_{i-1}}) +\\
+(-1)^p \left(\sum_{\sigma_i,\tau_i=0,1} \prod_{i=1}^{2N_p+p} e^{i\ell(\xi_i+3)(k_{\sigma_i}-k_{\tau_{i-1}})/2} \; 
\tilde{r}^{\sigma_i}(\xi_i,k_{\sigma_i})\; \tilde{r}^{\tau_{i-1}}(\xi_i,k_{\tau_{i-1}})\right). 
\end{multline}
The variables $\sigma_i,\tau_i=0,1$ have the same meaning as in~\eqref{eq:ugly}. The first term 
in~\eqref{eq:Mtpapp} corresponds to the choice $\eta_i=\xi_i+1$ for any $i$, and the second one 
is for $\eta_i=\xi_i+3$ (cf.~\eqref{eq:prodapp}). Now, the first term in~\eqref{eq:Mtpapp} is the 
same as in Ref.~\cite{alba2021unbounded}. The second one has the same structure, the only difference 
being that one has to replace $\xi_i\to\xi_1+2$. 
Following Ref.~\cite{alba2021unbounded}, we obtain 
\begin{multline}
\label{eq:Gcorr}
\widehat M_p^t=2^{2N_p+p}\int_{-k_F}^{k_F} \frac{dk}{2\pi}\Big[\ell-\frac{1}{2}\min(|v_k|t,\ell)+\frac{1}{2}(1-|a(k)|^2)^{2N_p+p}
\min(|v_k|t,\ell)\Big]+\\
+(-1)^p\Big[\ell-\frac{1}{2}\max(0,\min(|v_k|t-\ell,\ell))+\frac{1}{2}(1-|a(k)|^2)^{2N_p+p}\max(0,\min(|v_k|t-\ell,\ell))\Big]
\end{multline}
where to derive the second term we used that 
\begin{equation}
\int_{-1}^{1} d\xi \Theta(-\ell(\xi+3)/2+t|v_k|)=2\max(0,\min(|v_k|t/\ell-1,1)). 
\end{equation}

Let us now consider the effect of the term $\delta_{x,y}\sigma_z$ (cf.~\eqref{eq:Gammapm}). 
The idea is that one can replace any occurrence of a $G$ matrix in the definitions of $\Gamma^\pm$ with 
$\delta_{x,y}\sigma_z$. The result is 
\begin{multline}
\label{eq:moments}
\widehat M_p(\{n_j\})=\sum_{i=0}^{p}\binom{p}{i}  2^{p-i} (-1)^i 
\sum_{j_1=0}^{2n_1} \binom{2n_1}{j_1} 2^{2n_1-j_1} (-1)^{j_1}\cdots\sum_{j_p=0}^{2n_p} \binom{2n_p}{j_p} 2^{2n_p-j_p} (-1)^{j_p} \\
\int_{-k_F}^{k_F} \frac{dk}{2\pi}\Big[\ell-\frac{1}{2}\min(|v_k|t,\ell)+\frac{1}{2}(1-|a(k)|^2)^{p-i+\sum_{l=1}^p (2n_l-j_l)}\min(|v_k|t,\ell)\Big]\\
+(-1)^p\Big[\ell-\frac{1}{2}\max(0,\min(|v_k|t-\ell,\ell))+\frac{1}{2}(1-|a(k)|^2)^{p-i+\sum_{l=1}^p(2n_l-j_l)}\max(0,\min(|v_k|t-\ell,\ell))\Big]. 
\end{multline}
The sums and the combinatorial factors in the first row 
in~\eqref{eq:moments} account for all the possible insertions of $\delta_{x,y}\sigma_z$. 
By comparing~\eqref{eq:Gcorr} with~\eqref{eq:moments} it is clear that the effect of replacing the 
$G$ matrix with the term $\delta_{x,y}\sigma_z$ is an overall minus sign, 
a factor $1/2$ for each insertion, and 
a lowering of the exponent of the terms $(1-|a(k)|)^2$. 
Notice that the case in which we replace all the 
$\Gamma^\pm$s with $\delta_{x,y}\sigma_z$ has to be 
treated separately, as we will discuss in the following. 
The sums in~\eqref{eq:moments} can be performed exactly, yielding 
\begin{multline}
\label{eq:moments-1}
\widehat M_p=\int_{-k_F}^{k_F} \frac{dk}{2\pi}\left[\frac{\pi}{k_F}\ell-\frac{1}{2}\min(|v_k|t,\ell)+
\frac{1}{2}(1-2|a(k)|^2)^{p+N_p}\min(|v_k|t,\ell)\right]+\\
+(-1)^p\left[\frac{\pi}{k_F}\ell-\frac{1}{2}\max(0,\min(|v_k|t-\ell,\ell))+
\frac{1}{2}(1-|a(k)|^2)^{p+N_p}\max(0,\min(|v_k|t-\ell,\ell))\right].
\end{multline}
The term $\pi/k_F\ell$ in the square brackets in~\eqref{eq:moments-1} corresponds to replacing all 
the $\Gamma^\pm$ with $\delta_{x,y}\sigma_z$. 

To compute the negativity (cf.~\eqref{eq:neg-def}), we have to evaluate traces of powers of 
$P^{-1}(\Gamma^+ + \Gamma^-)$, where $P=\mathds{1} + \Gamma^+ \Gamma^-$. 
We proceed as in~\ref{sec:app-1}, rewriting 
$P^{-1}$ as 
\begin{equation}
P^{-1}=\sum_{n=0}^{\infty} (-1)^n (\Gamma^+ \Gamma^-)^n. 
\end{equation}
After using~\eqref{eq:moments-1}, and after summing over the 
variables $n_j\in[0,\infty)$ in~\eqref{eq:moments-1}, we obtain 
\begin{multline}
\label{eq:res-1}
\mathrm{Tr}\left[(P^{-1}(\Gamma^+ + \Gamma^-))^p\right]=\\
\int_{-k_F}^{k_F} \frac{dk}{2\pi}\left[\frac{\pi}{k_F}\ell-\frac{1}{2}\min(|v_k|t,\ell)+
\frac{1}{2}\left(\frac{2(1-2|a(k)|^2)}{1+(1-2|a(k)|^2)^2}\right)^p\min(|v_k|t,\ell)\right]\\
+(-1)^p\left[\frac{\pi}{k_F}\ell-\frac{1}{2}\max(0,\min(|v_k|t-\ell,\ell))+
\frac{1}{2}\left(\frac{2(1-2|a(k)|^2)}{1+(1-2|a(k)|^2)^2}\right)^p\max(0,\min(|v_k|t-\ell,\ell))\right],
\end{multline}
where we have used the regularisation $\sum_{n=0}^{\infty} (-1)^n=1/2$. It is straightforward to 
extend the result~\eqref{eq:res-1} to the trace of arbitrary functions of $P^{-1}(\Gamma^++\Gamma^-)$. 
For an analytic function $\mathcal{F}$, one finally obtains:
\begin{multline}
\mathrm{Tr}\left[\mathcal{F}(P^{-1}(\Gamma^+ + \Gamma^-))\right]=\\
\int_{-k_F}^{k_F} \frac{dk}{2\pi}\Big[\mathcal{F}(1)\Big(\frac{\pi}{k_F}\ell-\frac{1}{2}\min(|v_k|t,\ell)\Big)+
\frac{1}{2}\mathcal{F}\left(\frac{2(1-2|a(k)|^2)}{1+(1-2|a(k)|^2)^2}\right)\min(|v_k|t,\ell)\\
+\mathcal{F}(-1)\left(\frac{\pi}{k_F}\ell-\frac{1}{2}\max(0,\min(|v_k|t-\ell,\ell))\right)+
\frac{1}{2}\mathcal{F}\left(-\frac{2(1-2|a(k)|^2)}{1+(1-2|a(k)|^2)^2}\right)\max(0,\min(|v_k|t-\ell,\ell))\Big].
\end{multline}
As it is clear from~\eqref{eq:neg-def}, to evaluate the negativity, 
we have to use the function 
\begin{equation}
\mathcal{F}(x)=\ln\left(\sqrt{\frac{1+x}{2}}+\sqrt{\frac{1-x}{2}}\right),
\end{equation}
which is even and for which $\mathcal{F}(\pm1)=0$. We obtain the quite simple 
formula 
\begin{equation}
\label{eq:first}
\mathrm{Tr}\left[\mathcal{F}(P^{-1}(\Gamma^+ + \Gamma^-))\right]=
\int_{-k_F}^{k_F} \frac{dk}{2\pi} \mathcal{F}\left(\frac{2(1-2|a(k)|^2)}{1+(1-2|a(k)|^2)^2}\right)
\min(|v_k|t/2,\ell). 
\end{equation}
To obtain the negativity, we need the second term in~\eqref{eq:neg-def}. This is obtained 
from the results of Ref.~\cite{alba2021unbounded} as 
\begin{equation}
\label{eq:second}
\begin{gathered}
\mathrm{Tr}[\widetilde{\mathcal{F}}(G)]=\int_{-k_F}^{k_F} \frac{dk}{2\pi} \widetilde{\mathcal{F}}
(1-|a(k)|^2)\min(|v_k|t/2,\ell),
\end{gathered}
\end{equation}
with
\begin{equation}
\widetilde{\mathcal{F}}(x):=\ln(\sqrt{x^2+(1-x)^2}). 
\end{equation}
In~\eqref{eq:second} $G$ is the correlation matrix~\eqref{eq:Gxy} restricted to $A_1\cup A_2$. 
it is straightforward to show that the sum of~\eqref{eq:first} and 
\eqref{eq:second} vanishes.

%##########################
\section{Single fermion with a localized loss}
\label{sec:app-3}

Here we study the dynamics of a single fermion with 
momentum $k$ in the presence of a localized fermionic loss 
(see Ref.~\cite{burke2020non} for a similar setup). 
The system is a chain of $L$ sites. We work on a finite lattice,  to  have 
normalized states in what follows. 
The generic initial condition for a single fermion in a pure state  
$\rho(0)$ is: 
\begin{equation}
\rho(0)= \sum_{n,m}\psi_n (0) c^\dagger_n |0\rangle\langle0|c_m \psi^*_m(0).
\end{equation}
Here $c^\dagger_n$ are fermionic creation operators, $|0\rangle$ is the 
fermion vacuum, and $\psi_n(0)$ the wavefunction amplitudes.
The time evolution~\eqref{eq:lin} will transform the state into a 
mixed state $\rho(t)$ as 
\begin{equation}
\label{eq:rho-t}
\rho(t)= \sum_{n,m}\psi_n (t) c^\dagger_n |0\rangle\langle0|c_m \psi^*_m(t) + 
 C(t) |0\rangle \langle 0|, 
\end{equation}
where $C(t)$ is a parameter representing the probability of having lost the fermion at time $t$. 
The amplitudes $\psi_n(t)$ satisfy the system of equations 
\begin{equation}
\label{eq:psin-evol}
\frac{d \psi_n}{dt}= i (\psi_{n+1}+\psi_{n-1})-\frac{\gamma}{2}\delta_{n,0} \psi_0. 
\end{equation}
If we start from a plane wave $\psi_n(0)=L^{-1/2}e^{ikn}$, in the hydrodynamic limit 
$n,t\to\infty$ the solution of~\eqref{eq:psin-evol} is (see~\cite{alba2022noninteracting}) 
\begin{equation}
\label{eq:psi-t}
\psi_n(t)=\frac{1}{\sqrt{L}}e^{-i\varepsilon_k t}(e^{ikn}+r(k) \Theta(|v_k|t-|n|) e^{i|kn|}), 
\end{equation}
where $\varepsilon_k$ is the same energy dispersion in~\eqref{eq:vk} and 
$v_k$ is the fermion velocity. 
In the following, we consider as initial condition for our fermion the symmetric wavefunction $\psi_n(0)=(e^{ikn}+e^{-ikn})/\sqrt{2}$; by exploiting the linearity of~\eqref{eq:psin-evol}, it is immediate to obtain from~\eqref{eq:psi-t} the expression for $\psi_n(t)$. 

Let us now compute the fermionic negativity between two intervals of 
equal length $\ell$, $A=[-\ell,-1]$ and  $B=[1,\ell]$. 
The reduced density matrix of the two subsystems is obtained, by tracing over 
the rest of the chain, as 
\begin{equation}
	\rho_{AB}(t)=\sum_{|n|,|m|=1}^{\ell} \psi_n(t) c^\dagger_n |0\rangle \langle 0| c_m \psi^*_m(t) + M(t) |0\rangle \langle 0| := |\psi(t)\rangle\langle\psi(t)| + M(t)  |0\rangle \langle 0|, 
\end{equation}
where now $|0\rangle$ is the vacuum of fermions in region $A\cup B$ and $M(t)$ the probability for the fermion being 
absorbed or being outside region $A$. 
The state $|\psi(t)\rangle$ is clearly non-normalized. It is convenient to rewrite the previous equation as:
\begin{equation}
\rho_{AB}(t)= |\psi(t)|^2\; |\psi_N(t)\rangle\langle\psi_N(t)| + M(t)  |0\rangle \langle 0|,
\end{equation}
where the subscript $N$ means "normalized", and we have $\mathrm{Tr}[\rho_{AB}]=|\psi|^2+M=1$. 
For our initial condition, we have: 
\begin{equation}
|\psi|^2:=p^2=\sum_{|n|=1}^{\ell} 
\psi^*_n(t) \psi_n(t)= 
%\frac{1}{2L} 
%\sum_{|n|=1}^{l} (e^{ikm}+e^{-ikm}+2r(k)e^{ik|m|} \Theta(|v_k|t-|n|))\cdot\\
%\cdot(e^{ikm}+e^{-ikm}+2r(k)e^{-ik|m|} \Theta(|v_k|t-|n|))\simeq\\
%\simeq 
\frac{2\ell}{L} 
\left(1-|a(k)|^2 \min\left(\frac{|v_k|t}{\ell},1\right)\right) + O\left(\frac{1}{L}\right).
\end{equation}
To construct the partial transpose, we  
rewrite the density matrix in a basis that is a tensor product of a basis of $A$ and one of $B$. We define:
\begin{align}
& |\psi_A(t)\rangle = \sum_{n=-\ell}^{-1} 
\psi_n(t) c^\dagger_n |0\rangle = \frac{p}{\sqrt{2}} 
|\psi_{A,N}(t)\rangle;\\
& |\psi_B(t)\rangle = \sum_{n=1}^{\ell}
\psi_n(t) c^\dagger_n |0\rangle = \frac{p}{\sqrt{2}}  |\psi_{B,N}(t)\rangle.
\end{align}
Since $|\psi\rangle=|\psi_A\rangle + |\psi_B\rangle$, we can 
rewrite the density matrix in the basis 
$\{|0_A\rangle,|\psi_{A,N}\rangle\}\otimes\{|0_B\rangle,|\psi_{B,N}\rangle\}$ as:
\begin{equation}
\label{eq:single-state}
\rho_{AB}(t)=\begin{pmatrix}
M & 0    & 0 &    0\\
0 & p^2/2& p^2/2& 0\\
0 & p^2/2& p^2/2& 0\\
0 & 0    & 0    & 0\\
\end{pmatrix}
\end{equation}
And the partial transpose with respect to one of the two subsystems reads: 
\begin{equation}
\rho^{\mathrm{T_B}}_{AB}(t)=\begin{pmatrix}
M  & 0 & 0 & p^2/2\\
0  & p^2/2 & 0 & 0\\
0  & 0     & p^2/2&0\\
p^2/2&0&0&0\\
\end{pmatrix},
\end{equation}
with eigenvalues $\{p^2/2,p^2/2, (M+\sqrt{M^2+p^4})/2, 
(M-\sqrt{M^2+p^4})/2\}$. The negativity is, then:
\begin{equation}
\label{eq:single-negativity}
\mathcal{E}=\mathrm{Tr}(\ln|\rho_{AB}^{\top_B}|)=
\ln\left(1-M+\sqrt{M^2+(1-M)^2}\right)
\end{equation}
Formula~\eqref{eq:single-negativity} describes how the initial negativity ${\mathcal O}(\ell/L)$ decreases 
with time. Notice that~\eqref{eq:single-negativity} becomes the negativity content in~\eqref{eq:neg-den} after 
defining $1-M\to 2|a(k)|^2$. 

The single-fermion negativity sheds some light on the many-fermion case treated in the main text. Let us start 
from the state $b^\dagger_k b^\dagger_{-k}$. This state can be considered as a toy version of the Fermi sea. Also, 
the state $b^\dagger_k$ and $b^\dagger_{-k}$ can be interpreted as the transmitted and reflected fermions, respectively. 
One can rewrite $b_k^\dagger b^\dagger_{-k}$ as 
\begin{equation}
	b_k^\dagger b_{-k}^\dagger= \left(\frac{b_k^\dagger + b_{-k}^\dagger}{\sqrt{2}}\right) \left(\frac{b_k^\dagger - b_{-k}^\dagger}{\sqrt{2}}\right). 
\end{equation}
Now, it is easy to show that the antisymmetric combination $b^\dagger_k-b^\dagger_{-k}$  
anticommutes with the Lindblad operator $\sqrt{\gamma^-} c_0$, and does not evolve with time. On the other 
hand, the fermion in the symmetric combination can be absorbed at the impurity. After the fermion is absorbed, only 
the entangled antisymmetric combination remains.

\end{document}